\documentclass[useAMS]{mn2e}
\usepackage{epsfig}
\usepackage[dvips]{color}

\newcommand{\abs}[1]{\left\vert #1\right\vert}
\def\alt{\raise0.3ex\hbox{$\;<$\kern-0.75em\raise-1.1ex\hbox{$\sim\;$}}}
\def\agt{\raise0.3ex\hbox{$\;>$\kern-0.75em\raise-1.1ex\hbox{$\sim\;$}}}

\definecolor{Black}{named}{Black}
\definecolor{Red}{named}{Red}

\newcommand{\bw}{\begin{widetext}}
\newcommand{\ew}{\end{widetext}}
\def\d{{\rm d}}

\begin{document}

\title{Anisotropies in the Diffuse Gamma-Ray Background from Dark Matter with Fermi LAT: a closer look}
\author[A.~Cuoco, A.~Sellerholm, J.~Conrad \& S.~Hannestad]{
A.~Cuoco$^{1,2}$, A.~Sellerholm$^{1}$, J.~Conrad$^{1}$ and
S.~Hannestad$^{2}$
 \\
$^1$The Oskar Klein Centre for Cosmo Particle Physics, AlbaNova,
SE-106 91 Stockholm, Sweden \\
$^2$Department of Physics and Astronomy, University of Aarhus, Ny
Munkegade, Bygn. 1520 8000 Aarhus Denmark }

\maketitle

\date{\today}

\begin{abstract}
We perform a detailed study of the sensitivity to the anisotropies
related to Dark Matter (DM) annihilation in the Isotropic Gamma-Ray
Background (IGRB) as measured by the Fermi Large Area Telescope (Fermi-LAT).
For the first time, we take into account the
effects of the Galactic foregrounds and use a realistic
representation of the Fermi-LAT. We
implement an analysis pipeline which simulates Fermi-LAT data sets
starting from model maps of the Galactic foregrounds, the Fermi
resolved point sources, the extra-galactic diffuse emission and the
signal from DM annihilation. The effects of the detector are taken
into account by convolving the model maps with the Fermi-LAT
instrumental response. We then use the angular power spectrum to
characterize the anisotropy properties of the simulated data and to
study the sensitivity to DM. We consider DM anisotropies of
extra-galactic origin and of Galactic origin (which can be generated
through annihilation in the Milky Way sub-structures) as opposed to
a background of anisotropies generated by sources of astrophysical
origin, blazars for example. We find that with statistics from 5 years of observation
Fermi is sensitive to a DM contribution at the level of 1\%-10\% of
the measured IGRB depending on the DM mass $m_\chi$ and annihilation
mode. In terms of the thermally averaged cross section $\left<
\sigma_{\rm{A}}v\right>$, this corresponds to
$\sim 10^{-25}$cm$^{3}$s$^{-1}$, i.e. slightly above  the typical
expectations for a thermal relic, for low values of the DM mass
$m_\chi\alt 100$ GeV. The anisotropy method for DM searches has a
sensitivity comparable to the usual methods based only on the energy spectrum
and thus constitutes an independent and complementary piece of
information in the DM puzzle.
\end{abstract}


\section{Introduction}

Combined studies of the  cosmic microwave background  radiation,
supernova cosmology and large Galaxy redshift surveys provide
nowadays compelling evidence of the existence of non-baryonic Dark
Matter (DM) \cite{Komatsu:2008hk,Tegmark:2006az}. Consistency with
the observed structure of the Universe, especially at large
(above galactic) scales favors Cold Dark Matter (CDM), i.e. the particles
constituting the cosmologically required dark matter have to be
moving non-relativistically. This property is always fulfilled by
particles of mass in the GeV to TeV region that interact with the
weak interaction strength, i.e. weakly interacting massive particles
(WIMPs). WIMPs can annihilate or decay to detectable standard model
particles, in particular gamma-rays, giving rise to the possibility
of ``indirect detection''
\cite{Bertone:2004pz,Jungman:1995df,Bergstrom:2000pn,Baltz:2008wd}.
A  standard approach is to look for spectral signatures in the
energy spectrum of gamma-rays from celestial objects,
either targeting the continuum spectrum produced by decay of secondary pions or the more feeble signature of a line due to DM annihilating directly into photons.

Another possible experimental signature is given by differences in
the spatial distribution of gamma-ray signals induced by DM as
compared to conventional astrophysical sources. In particular, the
small scale fluctuations in an otherwise
isotropic gamma-ray background (IGRB)
might be different in the case of annihilating DM, since its flux scales
as the squared density. After the pioneering study of Ando and
Komatsu~\cite{Ando:2005xg} , followed by two other more detailed
works \cite{Ando:2006mt,Ando:2006cr}, the issue of DM annihilation
and anisotropies in the isotropic gamma background has raised
increasing interest. In connection to the extra-galactic gamma-ray background,  anisotropy has  been further studied in
\cite{Cuoco:2006tr,Cuoco:2007sh,Cuoco:2008zz} while predictions from the
Millenium II simulation have recently been described in 
Zavala et al. (2009). It has also been realized that Galactic
substructures can produce an almost isotropic gamma background  with
similar fluctuation properties as the extragalactic background and
with promising chances of detection
\cite{SiegalGaskins:2008ge,Ando:2009fp,SiegalGaskins:2009ux,Hensley:2009gh};
see also \cite{Pieri:2007ir,Pieri:2009je}). Other works have
investigated the anisotropy pattern resulting from both the Galactic
and extragalactic contribution \cite{Fornasa:2009qh,Hooper:2007be,Ibarra:2009nw}
or from the the DM annihilation around intermediate mass black holes
\cite{Taoso:2008qz}. Anisotropies from DM annihilation appearing in
the radio sky have been  investigated in \cite{Zhang:2008rs}.
Finally,-- apart from the power spectrum -- an interesting approach is
to investigate the probability distribution of the fluctuations in
order to detect possible departures from Gaussianity (or, rather,
from Poisson statistics, in the common situation of low number of
counts), thereby providing another handle on the separation of
conventional astrophysical and DM contribution
\cite{Lee:2008fm,Dodelson:2009ih}. Anisotropy properties of
astrophysical processes other than DM contributing to the isotropic
gamma background include resolved point sources \cite{Ando:2006mt},
galaxy clusters and shocks \cite{Miniati:2007ke,Keshet:2002sw} and
emission from normal galaxies \cite{Ando:2009nk}.

The Fermi Large Area Telescope (Fermi-LAT), which was launched
June 11, 2008, is the most suitable tool for these kind of studies
due to its large field of view ($\sim$ 2.4 sr), energy coverage from
100 MeV to 300 GeV \cite{Atwood:2009ez} and very good
angular resolution. We will thus focus in the following on the prospect to detect DM anisotropies with Fermi-LAT.

All of the above studies assume either that the impact of Galactic
foregrounds on the DM induced anisotropies is negligible or that the
Galactic foregrounds can be cleaned at a suitable level to allow DM
anisotropies studies.  Further, the effects of the Fermi-LAT
detector are generally taken into account only in an approximate
way, though a detailed representation of the Fermi instrument is
publicly available.

The aim of this work is to address the above issues in detail. In
particular our approach will be the following: we include the
Galactic foregrounds explicitly using a model based on the GALPROP
code \cite{Strong:1998pw,Moskalenko:1997gh,Strong:2004de} and we then use this model to build masks to select the most
promising regions in the sky suitable for the analysis of the
anisotropies. In these regions we study the anisotropies in the
scenario in which a DM contribution is present over the Galactic
foregrounds and the extra-galactic emission from blazars compared to
the case without DM and we derive the sensitivity to the DM signal
employing anisotropy studies with Fermi-LAT data.
Both the cases of Galactic DM in substructures and extra-galactic DM
are considered separately. The effect of the Fermi-LAT response,
including the effects of the point spread function, the effective area, as well as
non-uniformities in exposure, are included in all calculations at a
realistic level by using the instrument response functions provided by
the Fermi Science Tools.

This approach is suitable for a sensitivity study like the one
presented here, and does not aim to fully reproduce a real data
analysis for which foreground cleaning (not performed in this study) most likely will be required.
One of the findings of our analysis is indeed that even with a very
generous masking, foregrounds still have a sizeable contribution to
the overall anisotropies and represent an important limitation to DM
sensitivity. This implies that foreground cleaning will be a
crucial step in the data analysis pipeline. On the other hand, if
cleaning can be achieved in an effective way, this also means that
the sensitivity to DM can be further improved with respect to the
results derived in this paper. Our results can thus be regarded as
conservative from this point of view. Foreground cleaning will be
the subject of forthcoming investigations.

The paper is structured as follows. In section II we introduce and
discuss the various components contributing to the Galactic and extragalactic gamma-ray emission. In sections III and IV we
review their anisotropy properties and we describe the tools and
methods for their analysis. In section V we  briefly describe the
Fermi-LAT instrument and the characteristics relevant for our
analysis. In section VI we discuss the results of our data
simulation pipeline. Finally, in section VII we evaluate Fermi's
sensitivity to distinguish a scenario in which the diffuse gamma-ray
emission has a DM contribution (either extragalactic or galactic in
origin) with respect to the case in which DM is absent. Conclusions
are presented in section VIII.

\begin{figure*}
\vspace{-0.0pc}
\begin{center}
\vspace{-0.5pc}\includegraphics[width=1.90\columnwidth,angle=0]{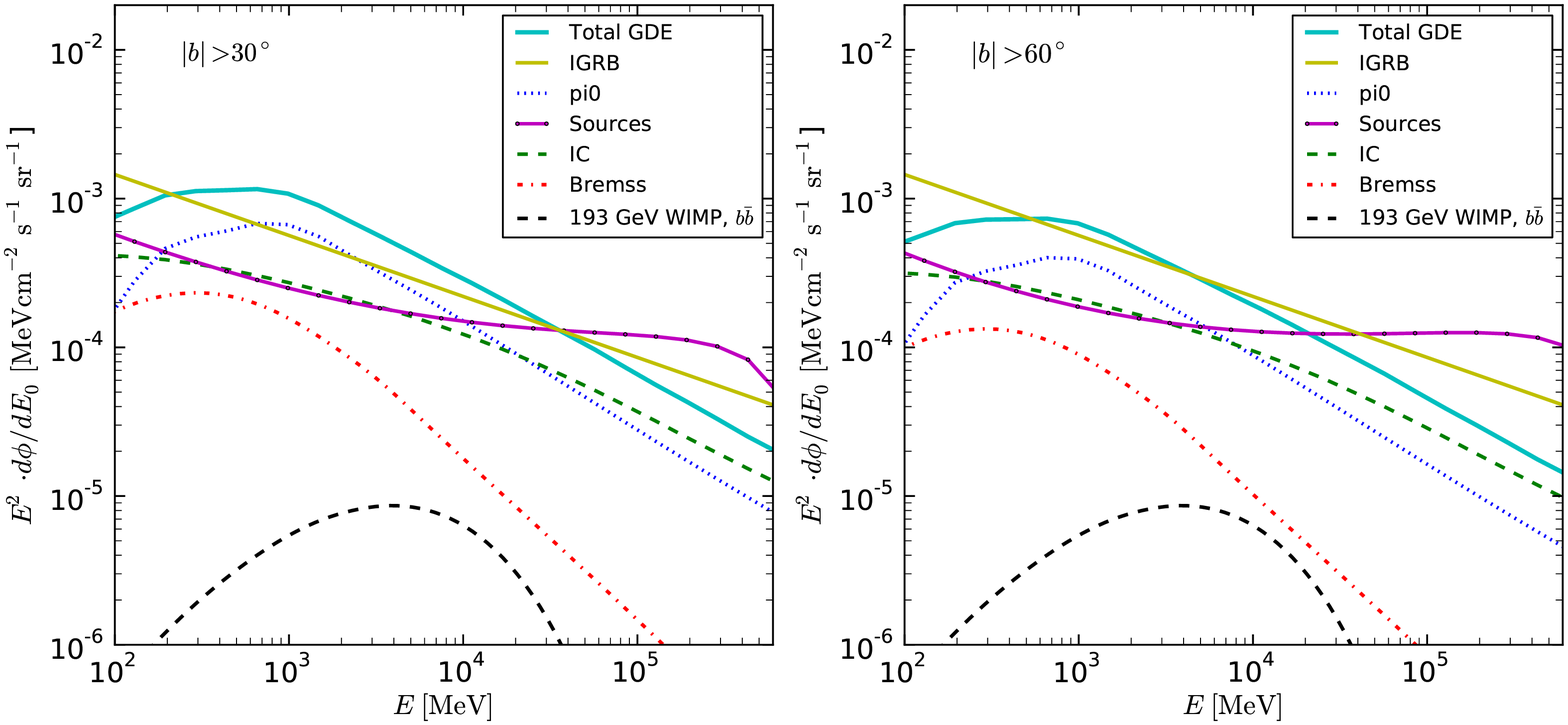}
\end{center}
\vspace{-1.0pc} \caption{Energy spectrum of the various components
contributing to the gamma sky above $30^\circ$ (left) and above
$60^\circ$ (right) galactic latitude. The isotropic component is the
determination from Abdo et al. (2009b). Also shown is an example of
extragalactic WIMP emission normalized with the model of Ullio et
al. (2002) (with NFW halo profile and no halo substructures). The
spectrum of the resolved point sources is calculated from the Fermi
11-month source catalogue (Ballet 2009).
 \label{fig:diffuse}}
\end{figure*}

\section{Gamma-ray sky Components}

\subsection{Isotropic Gamma-Ray Background}
In the following, for clarity,  we will refer to the isotropic
experimentally detected emission as the Isotropic Gamma-Ray
Background (IGRB) given that in principle its origin may be not
fully extra-galactic. Indeed, we will also consider the possibility of a nearly isotropic emission from DM annihilation in the Milky Way
substructures. We will use the term Extra-Galactic Gamma
Background (EGB) when referring to predictions from \emph{models} of
extra-galactic emission either from DM annihilation or from
astrophysical processes.

A first  detection of the IGRB in the GeV range was
reported by the EGRET experiment
\cite{Sreekumar:1997un} (see also the analysis  by Strong,
Moskalenko and Reimer \cite{Strong:2004ry} ). 
The recent measurement from The Fermi collaboration  \cite{Ackermann:prl} finds 
that the IGRB energy spectrum has an almost perfect power law with slope
$s\approx -2.4$ which is approximately what is expected from the
contribution of unresolved blazars thus suggesting an extragalactic
origin of the IGRB. 

A (possibly incomplete) list of contributions to the  IGRB  is given
by the emission in astrophysical processes from Blazars, AGNs,
Starburst Galaxies, Star Forming Galaxies, Galaxy clusters, Clusters
Shocks, Gamma-Ray Bursts
\cite{Stecker:1996ma,Pavlidou:2002va,Gabici:2002fg,Totani:1998xc}. A
contribution is also expected from ultra high energy photons and
electrons cascading to GeV-TeV energies through interactions with
the CMB or the Infrared-Optical Background \cite{Kalashev:2007sn}.
Each of these source classes is expected to produce anisotropies although
all them should more or less trace the filamentary pattern of the
cosmological Large Scale Structures. Anisotropies will be discussed in
more detail in the next two sections.

\subsection{DM Annihilation}
Apart from the astrophysical contributions, an exotic component
arising from WIMP annihilation could also be  expected. In the case
of uniformly distributed DM, the annihilation flux for a thermal
relic cross section ( $\left< \sigma_{\rm{A}}v\right> \sim
3\times10^{-26}$cm$^{3}$s$^{-1}$ ) is many orders magnitude below the IGRB. However,
given that the DM distribution on cosmological scales could be very
clumpy, the enhancement factor (denoted as $\Delta^2(z)$) due to the
quadratic dependence of the annihilation signal on the DM density
has to be taken into account. This enhancement is typically several
orders of magnitude, thus boosting the DM signal to a level
comparable to the IGRB
\cite{Bergstrom:2001jj,Ullio:2002pj,Taylor:2002zd}. The exact boost
depends on the details of the modeling of the DM clustering, the DM
halo profile and the structure formation history. We fix the
absolute normalization of the DM spectra following the DM haloes
clustering model described in Ullio et al.(2002). Two versions of
the model are employed: a conservative version assuming a
Navarro-Frenk-White (NFW) profile \cite{Navarro:1995iw} for the
haloes and no substructures, which gives a low DM normalization, and a
more optimistic version where the haloes still follow a NFW profile
but the DM signal is enhanced by the presence of substructures in
the DM haloes. It should be noted that the clustering properties of
the DM haloes are still very uncertain, especially at small scales
(approximately below the typical galactic halo scale), and even
larger normalizations are possible. For example recent results from
the Millenium-II simulation \cite{Zavala:2009zr} indicate that,
depending on the extrapolation employed for the low mass haloes (non
resolved in the simulation), an order of magnitude enhancement with
respect to the above ``optimistic'' case can be achieved. Overall, 
\mbox{Zavala et al. (2009)} place the possible values of the boost $\Delta^2(z$$=$$1)$ in the
range $10^4$ to $10^7$ (see also \cite{Abdo:2010dk}, in particular Fig.1).
\subsection{Galactic Foregrounds}

The main contribution to the Galactic foregrounds is given by the
decay of pions produced in the interaction of Cosmic Ray (CR)
nucleons, by the inverse Compton (IC) scattering of CR electrons on
the interstellar radiation field (ISRF) and by 
bremsstrahlung of CR electrons on the interstellar gas. A further contribution from
pulsars has also been considered recently
\cite{FaucherGiguere:2009df}. According to this analysis, the
contribution from Galactic millisecond pulsars can be quite
isotropic (i.e. can contribute to high Galactic latitudes) and
constitute a large fraction of the IGRB. However, lacking more precise studies of the subject, we will neglect
this contribution for the present analysis.

The normalization of the Galactic foregrounds is clearly important
to identify the regions where the IGRB is dominant, i.e. the regions most
promising for the anisotropy analysis. We employ the so called ``conventional
model'' \cite{Strong:1998pw,Moskalenko:1997gh,Strong:2004de} which is
derived under the assumption that the CR electron and nucleon spectra in the
galaxy can be normalized by the local (solar system) measurements.
This model represents a nice
fit of the Fermi data, at least at Galactic latitudes $\abs{b}>10^\circ$, where the
emission is mostly local in origin (coming within a few kpc's of the
solar position) \cite{Porter:2009sg,Abdo:2009mr,Porter:TeVPa}. 
{We will, indeed,  focus on this region of the sky in the following. The  
Galactic center region and the Galactic plane are anyway 
strongly dominated by the Galactic emission and they are thus not suitable 
for the anisotropy analysis. A description of the diffuse emission in terms 
of the conventional model is thus accurate enough for the present analysis.}

\subsection{Resolved Point Sources}
The number of resolved point sources has grown considerably in the
Fermi era relative to  the roughly 270 in the 3rd EGRET catalogue \cite{1999yCat..21230079H}.
The first year Fermi catalogue includes  1451 sources with a
significance larger than about 4 sigma \cite{Collaboration:2010ru}.
We include explicitly these sources in our analysis. As will be clear in the
following, it is important to include and, in addition, to model this
component correctly since it is a considerable source of anisotropy which
needs to be reduced to gain sensitivity to the anisotropies of the
diffuse component.

{
While we use the first year catalogue, in the following we produce a forecast over 5 years of data taking, thus should in principle be using an associated 5 years catalogue. The effect of using a five year catalogue, however, is expected to be small: the masked area will increase, but on the other hand the Poisson noise will be reduced. Both effects are small since increasing the data taking time for the catalogue will add sources on the faint end, requiring small masking area and producing a small decrease in Poisson noise (since the brightest sources are already removed). On the other hand, modeling the 5 year catalogue requires an assumption on the nature of the unresolved sources, introducing a model uncertainty. We therefore decided to use the 1 year catalogue only.
}

\subsection{Unresolved Point Sources}
Part of the IGRB is made of \emph{unresolved point sources}. It is
still a matter of debate if unresolved point sources can make up all
the IGRB or only part of it. This issue is likely to become clearer
when population studies of the Fermi \emph{resolved point sources}
become available. Present studies indicate that unresolved blazars can only account for at most 30 \% of the IGRB measured by the Fermi-LAT \cite{Collaboration:2010gq}.

Unresolved point sources can also be a further
source of anisotropy in the IGRB itself, producing a
Poisson noise--like contribution to the anisotropy spectrum. It is worthwhile to point out that  this Poisson noise is
different from the Poisson shot noise in the angular power spectrum
(see below and appendix A), which is related to the finite number of
counts. The contribution of unresolved point sources depends both on
the ability of the Fermi-LAT to  detect point sources and remove
them from the diffuse emission  and on the intrinsic number density
of the sources themselves. Both these terms, in turn, are quite
model dependent but they are expected to be reasonably constrained
from population studies of source classes with resolved members.

{
For the moment, for the sake of studying the importance of the effect, we consider, beside the model described in the next section,  a model fully made of unresolved point sources. For the Poisson noise we assume a value of $C_P=5.0 \times 10^{-5}$ which is in the middle between the estimate of $10^{-4}$ \cite{Ando:2006cr} (see their Fig.4) and the preliminary measurement of $10^{-5}$ from the Fermi collaboration \cite{SiegalGaskins:2010nh}. 

It is also worth noting that the Poisson noise of
unresolved sources is a time varying term slowly
decreasing in time. The brighter unresolved sources, in fact, get finally resolved, as
Fermi continues to collect more statistics, and once they are removed, and/or masked, the ``new'' remaining  unresolved signal has a slightly lower level of Poisson noise. A detailed study of this effect
requires extrapolation from population studies which we leave for future work. }

In Figure \ref{fig:diffuse} we show the energy spectrum of the
various Galactic and extra-galactic components for the model we
employ. In particular, we show the average spectrum in the regions
$\abs{b}>30^\circ$ and $\abs{b}>60^\circ$, where the IGRB has the
highest relative contribution. We show the power law fit to of the IGRB from  \cite{Ackermann:prl}. 
Also shown is an
example of extragalactic WIMP emission normalized to the model of
Ullio et al. (2002) (with NFW halo profile and no halo
substructures). The contribution from resolved point sources
detected by Fermi \cite{Abdo:2009mg},\cite{Ballet:Symposium} is also
present. It can be seen that, combined, these sources
constitute a significant fraction of the diffuse background. For
simplicity, the spectrum of each single source is modeled as a
simple power law and thus the behavior at high energies ($\agt10$
GeV) may have some biases with respect to the real spectrum.

\begin{figure}
\vspace{-1.0pc}
\begin{center}
\begin{tabular}{c}
\vspace{-0.5pc}\includegraphics[width=0.99\columnwidth,angle=0]{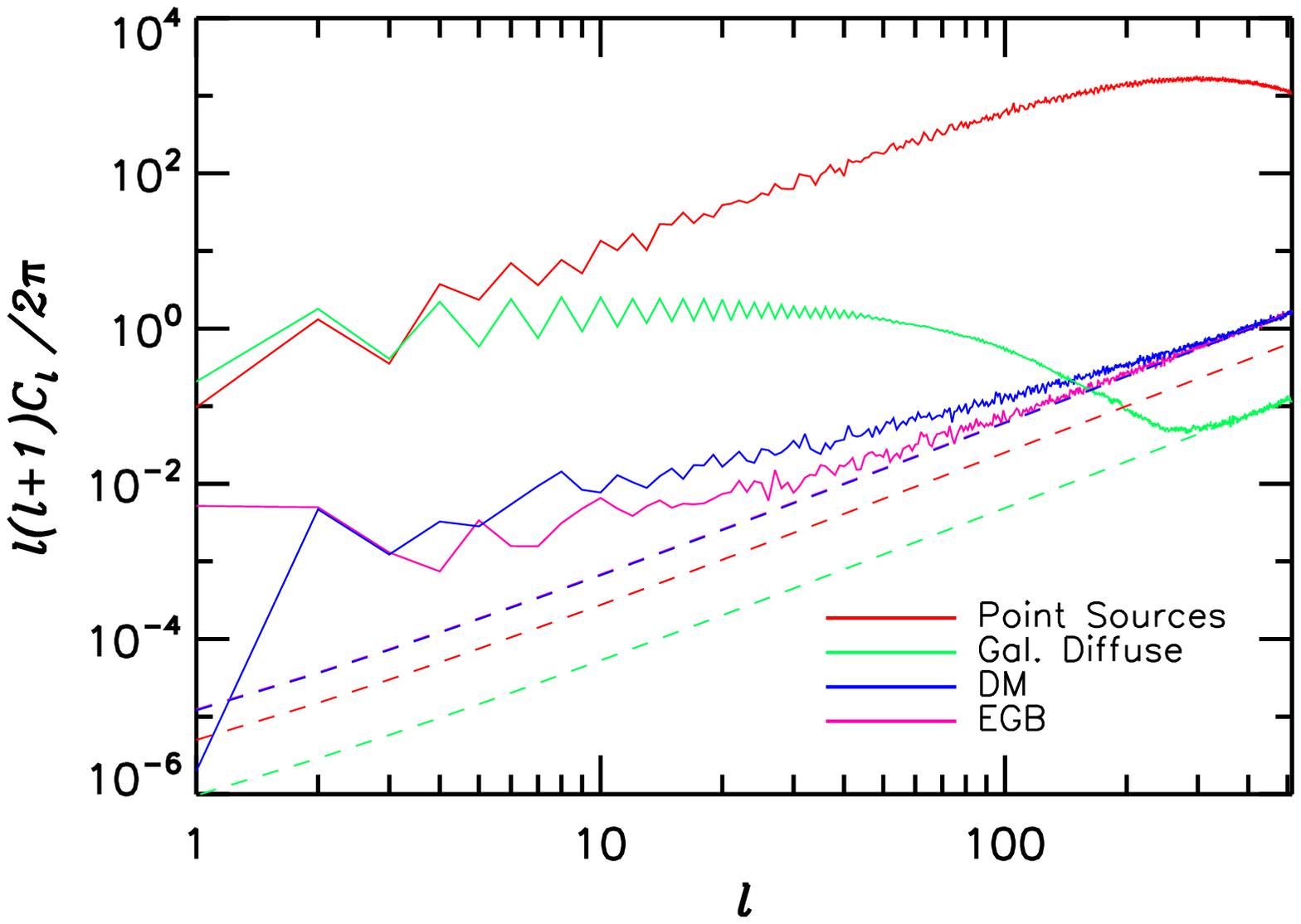}
\\
\vspace{-0.5pc}\includegraphics[width=0.99\columnwidth,angle=0]{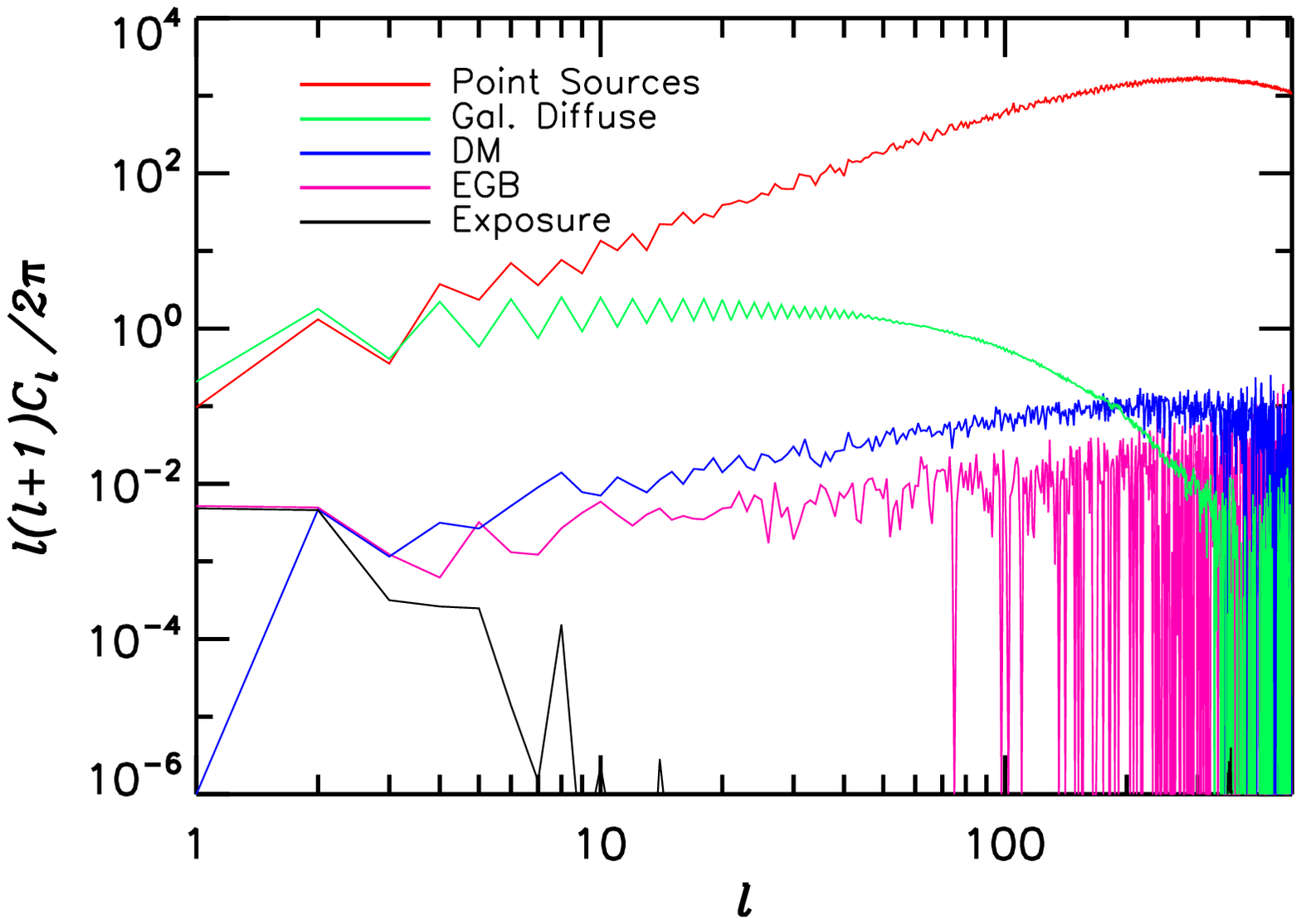}
\\
\vspace{-0.5pc}\includegraphics[width=0.99\columnwidth,angle=0]{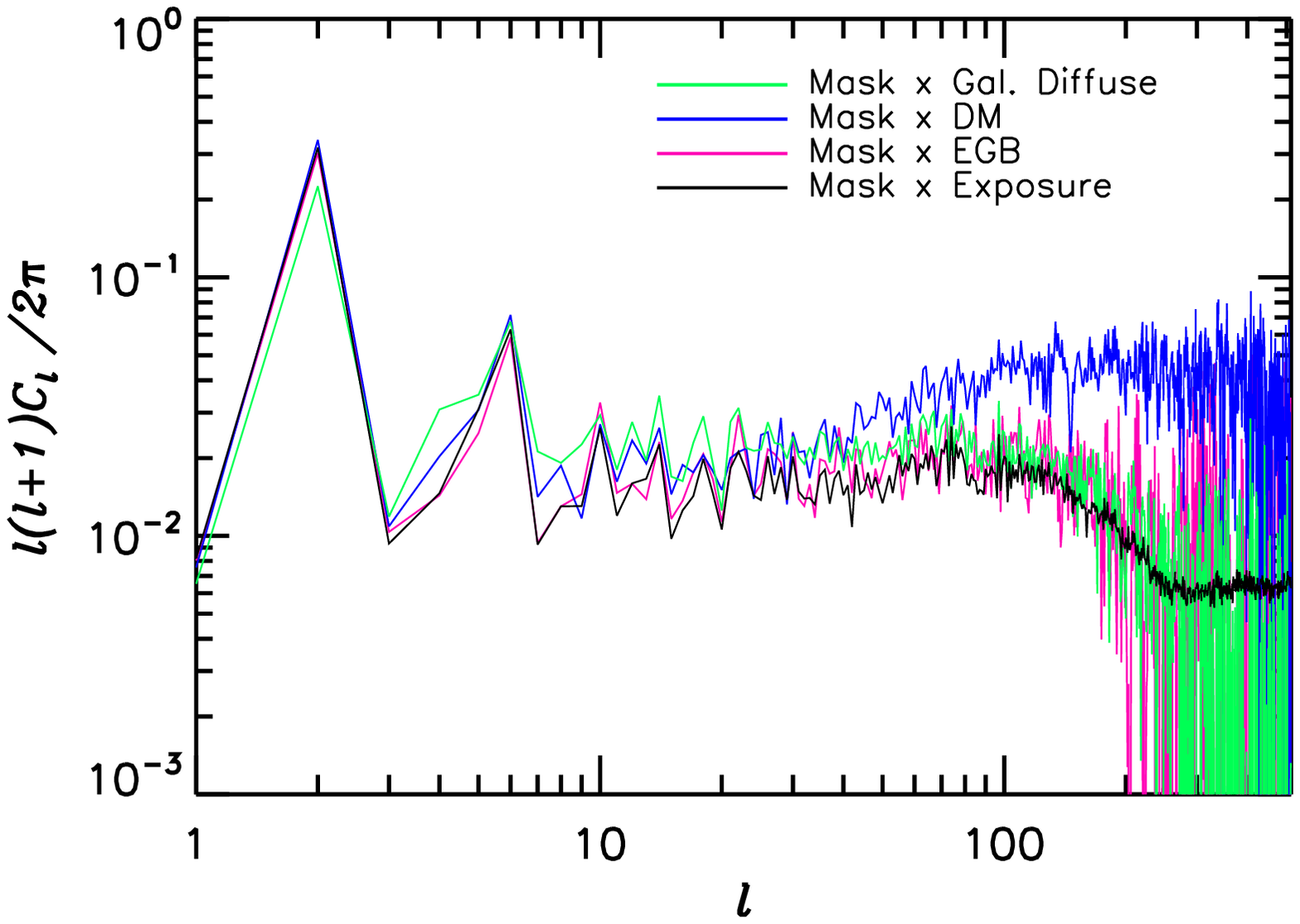}\\
\end{tabular}
\vspace{-0.5pc}
\end{center}
\hspace{-1.5pc} \caption{All sky angular spectra for $E>1$ GeV of the
employed models for point sources (red), Galactic foregrounds
(green), EGB (magenta) and EGB DM (blue) after convolution with the
Fermi-LAT Point Spread Function. For illustration, the flux of the DM
component is arbitrarily normalized to the level of the IGRB
 detected by Fermi-LAT, and 5 years of Fermi-LAT observations have been
assumed. The power spectrum of the exposure map is also shown
(black line). In the top panel the spectra before shot noise removal are shown 
(shot noise is represented by the dashed lines) . All the spectra
are calculated with HEALpix. The last panel shows the angular power
spectra after the application of  a suitable mask to cover the low latitude
Galactic foregrounds and the point sources. Since the mask is effective in 
suppressing the point sources signal the related power spectrum has been removed.
\label{fig:3}}
\end{figure}

\section{Anisotropies: General Considerations}
We will follow two approaches to quantify the anisotropies of all
the gamma-ray sky components. For the components for which maps as a
function of energy are available (i.e. for the Galactic foregrounds
and for the resolved point sources) we calculate the angular power
spectra directly from the maps using HEALpix \cite{Gorski:2004by}(see appendix A for
more details on HEALpix and the definition of power spectra). For
the remaining components (DM and astrophysical EGB) we model the
power spectra from theory. From the theoretical power spectra
we then also produce simulated sky-maps (under the assumption that the
fluctuations are gaussian). In this way we have effective sky maps
for all the components and we can combine them, convolve them
with the Fermi IRFs, simulate finite statistics of events, apply
masks, etc. The final maps will thus represent a simulated gamma-ray
sky as seen through the Fermi detector and where all
the components are entangled. {Power spectra are then recovered at
the end from the final maps using HEALpix.} In this way the
sensitivity to DM can be studied as described
in section \ref{sec_sense} by comparing the power spectra of the maps with and without the DM component.

Fig.\ref{fig:3} shows separately the all-sky angular spectra of the
various components: Galactic foregrounds, astrophysical EGB, WIMPs
EGB and point sources. The spectra are calculated from simulated
maps at $E>1$ GeV after convolving with the Fermi IRFs (see the
next sections). For illustrative purposes, the flux of the DM
component is arbitrarily normalized to the level of the IGRB detected
by Fermi and an observation time of 5 years is
assumed. It can be seen that the Galactic foreground anisotropies
have a decreasing spectrum in $l$ and dominate over the
extragalactic anisotropies (either astrophysical or WIMP) up to
$l\sim 200-300$. Luckily, however, this can be remedied using an
appropriate mask excluding the Galactic plane region where most of
the foregrounds are concentrated. In this way the foregrounds can be
reduced such as  to become subdominant at a multipole of $l\sim30$ as
shown in the lowest panel of fig.\ref{fig:3}. The spectrum of point
sources highly dominates both the foregrounds and the extragalactic
signals. This indicates that an efficient point source masking is
crucial to extract the interesting extragalactic signal. These steps
will be explored in the next sections. The figure also illustrates
the effect of the Poisson noise (first panel)  coming from finite
statistics present in the calculated raw power spectra. Customary
power spectra are plotted with the noise removed (second panel).

The following section describes in detail how the power
spectra of DM and astrophysical EGB are modeled and how the related
maps are simulated. The section can be skipped by the reader not
interested in these details.

\section{Dark Matter and Astrophysical Anisotropies}
\subsection{Modeling the EGB}
Apart the Poisson term coming from the unresolved point
sources,  the remaining source of anisotropies of the IGRB is given
by the anisotropic spatial distribution of the sources themselves.
To derive the anisotropy we will assume, as a reasonable first 
approximation, that the gamma ray sources are distributed as the
matter density of the universe $\rho(\vec{x})$, i.e. following the
cosmological Large Scale Structures (LSS). In principle $\rho_s$,
the density distribution of astrophysical sources, should be used
instead of $\rho$: $\rho_s$ in general exhibits a scale and time
dependent bias with respect to the matter density. However, specific
classes of astrophysical gamma-ray sources have different biases.
For example, blazars  are well known to concentrate at the center of clusters of
galaxies, thus presenting an over-bias with respect to galaxies at
high densities. On the other hand, galaxies and clusters of galaxies reasonably
trace the matter density, at least in the recent cosmic
epoch. The assumption $\rho_s=\rho$  is thus general
enough to approximately describe emission from astrophysical
sources.

Given these assumptions the extragalactic cosmic gamma-ray signal can
be written as \cite{Ullio:2002pj,Bergstrom:2001jj,Cuoco:2006tr}
\begin{equation}
I_{\gamma}(E_\gamma,\hat{n})\propto \int_0^\infty \!\!\!\d z\,
\frac{\rho(z,\hat{n},r(z))\,g[E_\gamma(1+z)]\,e^{-\tau(E_\gamma,z)}}{H(z)\,(1+z)^{3}}
\,,\label{intcosmo}
\end{equation}
where  $g(E)=dN_\gamma/dE$ is the photon spectrum of the sources,
$E_\gamma$ is the energy we observe today, $\rho(z,\hat{n},r)$ is
the matter density in the direction $\hat{n}$ at a comoving distance
$r$, and the redshift $z$ is used as time variable. The Hubble
expansion rate is related to its present $z\!=\!0$ value $H_0$
through the matter and cosmological constant energy densities as
$H(z)= H_0\sqrt{\Omega_M(1+z)^3+\Omega_\Lambda}$, and the reduced
Hubble expansion rate $h(z)$ is given by $H(z)=100 \ h(z)$ km/s/Mpc.
We will in the following use $\Omega_M = 0.3$, $\Omega_\Lambda = 0.7$
and $H_0 = 70$ km/s/Mpc. The quantity $\tau(E_\gamma,z)$ is the
optical depth of photons to absorptions via pair production (PP) on
the  Extra-galactic Background Light (EBL).  We use the
parametrization of $\tau(E_\gamma,z)$ from \cite{Stecker:2005qs} for
$0\!<\!z\!<\!5$, where the evolution of the EBL is included in the
calculation. The EBL is expected to be negligible at redshifts
larger than $z\approx 5$ corresponding to the peak of star
formation. Thus, gamma photons produced at earlier times experience
an undisturbed propagation until $z\approx 5$, while only in the
recent epoch they start to lose energy due to scattering on the
EBL. Correspondingly, we assume $\tau(E_\gamma,z)=\tau(E_\gamma,5)$
for $z>5$ (see also formula (A.6) in \cite{Cuoco:2006tr}).

In the case of cosmological DM annihilation, the resulting spatial
distribution of the gamma signal follows the square of the matter
distribution $\rho^2(\vec{x})$ through
\begin{equation}
I_{\chi}(E_\gamma,\hat{n}) = \frac{ <\!\!\!  \sigma_\chi v \!\!\!
>}{8\pi m^2_\chi}  \int_0^\infty \!\!\!\d z\,
\frac{\rho_\chi^2(z,\hat{n},r(z))\,g[E_\gamma(1+z)]\,e^{-\tau(E_\gamma,z)}}{H(z)\,(1+z)^{3}}
\, ,\label{intcosmoDM}
\end{equation}
which, taking the spatial average, reduces to the
expression
\begin{equation}
I_{\chi}(E_\gamma) = \frac{ <\!\!\!  \sigma_\chi v \!\!\!
> \bar{\rho_0}^2}{8\pi m^2_\chi}  \int_0^\infty \!\!\!\d z\,
\frac{\Delta^2(z)\,(1+z)^{3}\,g[E_\gamma(1+z)]\,e^{-\tau(E_\gamma,z)}}{H(z)}
\, ,\label{intcosmoDM}
\end{equation}
where $\Delta^2(z)$ is the clumpiness enhancement factor and
$\bar{\rho_0}$ is the average DM density at $z=0$.

We can write the astrophysical EGB and DM in a compact form: 
\begin{eqnarray}\label{gammaflux1}
I_{\gamma}(E_{1},E_{2},\hat{n}) &\propto&\! \int_0^\infty \!\!\!\d
z\, W_{\gamma}(E_{1},E_{2},z) \,
\rho (z,\hat{n})\,, \\
\label{gammaflux2} I_{\chi}(E_{1},E_{2},\hat{n}) &\propto&\!
\int_0^\infty \!\!\!\d z\, W_{\chi}(E_{1},E_{2},z) \, \rho^2
(z,\hat{n})\,,
\end{eqnarray}
where $W_{\gamma}(E_{1},E_{2},z)$ and $W_{\chi}(E_{1},E_{2},z)$ are
the astrophysical and DM window functions, which contain the
information about gamma-ray propagation, injection spectra and
cosmological effects
\begin{equation}
W(E_{1},E_{2},z)\!\equiv\!\int_{E_{1}}^{E_2}\!\!\!\!\!\!\!\d E\,
 \frac{ g[E(1+z)] \,(1+z)^{3\alpha-3}\!\!}{H(z)} \, e^{-\tau(E,z)}\!,
\label{gammawindow}
\end{equation}
where $\alpha=1,2$ applies in the astrophysical and DM cases,
respectively and $E_1$ and $E_2$ are the boundaries of the energy
bands considered. We are using the notation $\rho
(z,\hat{n},r(z))=(1\!+\!z)^3 \times \rho (z,\hat{n})$ to underline
that the window function is only dependent on the two variables,
direction and redshift, and to make the $(1\!+\!z)^3$ behavior of
the matter density explicit. In particular, choosing $E_1=E_{\rm
cut}$ and $E_2=+\infty$, when properly normalized $W(E_{\rm
cut},z)$ represents the probability of receiving a photon of
$E_\gamma> E_{\rm cut}$ emitted at a redshift $z$. This can be used
to define an effective horizon, $z_{\cal H}$, beyond which the
probability of receiving a photon is negligible. In practice, above
about 10 GeV, pair production attenuation sets the scale of the
horizon which is about $z_{\cal H}\sim 0.5$ and $z_{\cal H}\sim 2$
for the $\alpha=1,2$ cases respectively. In this case, DM annihilation
produces a much larger anisotropy than the astrophysical sources. In
contrast, below 10 GeV the horizon is set by cosmological effects
and $z_{\cal H}\sim 1$ and $z_{\cal H}\sim 10$ for the two cases. In
the low energy regime we thus see that the DM anisotropies are
averaged over a much larger horizon and the final signal can be
smoother than the astrophysical case. Further details are discussed
in \cite{Cuoco:2007sh}.

The injection energy spectrum for DM annihilation $g(E_\gamma)=
dN_\gamma/dE$, given the annihilation channel and the WIMP mass
$m_\chi$, is calculated with DarkSUSY~\cite{Gondolo:2004sc}, which, in
turn, uses a tabulation derived from simulation of the particle
processes performed with Pythia~\cite{Sjostrand:2007gs}. We will
consider in the following the $b\bar{b}$, $\tau^+\tau^-$ and
$\mu^+\mu^-$ annihilation channels.

\subsection{Angular Power Spectra}

Given the above expressions  we can write the angular power
spectra of the various dimensionless fluctuation fields \mbox{$\delta I/\!\!<\!\!I\!\!>$}
\begin{eqnarray}\label{angspectra}
  C^l_{\gamma} & = &  \int \frac{dr}{r^2} \; W_{\gamma}^2(r) \;
    P_{\rho}\left(k=\frac{l}{r},z(r)\right)\, , \\
  C^l_{{\chi}} & = &  \int \frac{dr}{r^2} \; W_{{\chi}}^2(r) \;
    P_{\rho^2}\left(k=\frac{l}{r},z(r)\right)\, ,
\end{eqnarray}
where we have used the Limber approximation \cite{Limber}, which is
accurate for all but the very lowest multipoles, while
$P_{\rho}\left(k,z(r)\right)$ and $P_{\rho^2}\left(k,z(r)\right)$
are 3D power spectra of the matter field and of its square. We are
also interested in the spectra of cross correlation among two
different energy bands, which can be written as
\begin{equation}\label{angcrossspectra}
  C^l_{12}  =   \int \frac{dr}{r^2} \; W_{1}(r)W_{2}(r) \;
    P_{i}\left(k=\frac{l}{r},z(r)\right)\, ,
 \end{equation}
where $i=\rho, \rho^2$ for the astrophysical and DM case. 

We take the 3D power spectra $P_{\rho}$ and $P_{\rho^2}$ from an N-body
simulation of Large Scale Structures formation \cite{Cuoco:2007sh}.
We note that $P_{\rho^2}$   has a  dependence on
the clustering properties of DM below galactic scales (see
\cite{Ando:2005xg} for an analytical derivation of $P_{\rho^2}$)
which are not resolved in the N-body simulation. The anisotropy
properties in the case of cosmological DM annihilation with
substructures are therefore not fully consistent with its energy
spectrum normalization, which does take into account the effects of
substructures. This has to be taken into account when interpreting the
sensitivities in the cosmological substructures model shown in the following.

{
As described in the previous section we will consider in the following two different anisotropy models for the astrophysical EGB in order to bracket our uncertainties and estimate the effect on the DM sensitivity. The first one is a model dominated by the correlation term of the power spectrum (Eq.\ref{angspectra}), while the second one is a model dominated by the Poisson term  $C_P$ from unresolved point sources. In this case the angular power spectrum is simply:
\begin{equation}\label{angcrossspectra}
  C^l_{\gamma} = C_P
 \end{equation}
with $C_P=5 \times 10^{-5}$.  }

\begin{figure}
\vspace{0.0pc}
\begin{center}
\vspace{-0.5pc}\includegraphics[width=0.99\columnwidth,angle=0]{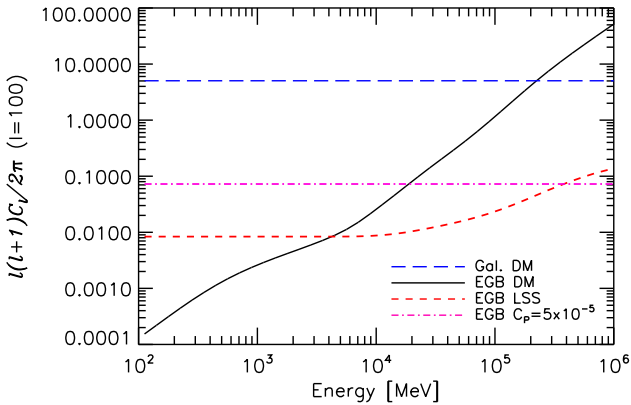}
\vspace{-2.5pc}
\end{center}
\hspace{-1.5pc} \caption{Energy dependence of the anisotropies for
our models of cosmological EGB DM, astrophysical EGB with and without a dominant Poisson term and DM in
Galactic substructures. The anisotropy curve for the DM refers to a $m_\chi$=1 TeV WIMP annihilating
into $b\bar{b}$.\label{fig:anieni}}
\end{figure}

Finally we will also consider the possibility that part of the IGRB
is produced by DM annihilation in the sub-structures of our galaxy
as found by recent N-body simulations
\cite{Diemand:2006ik,Springel:2008by,Springel:2008cc}. In this case,
compared to the cosmological annihilation scenario, the situation is
simpler since, in the absence of  redshift effects, the pattern of
anisotropies is independent of energy (and thus of  the particle
physics model). This can be seen more clearly in
Eq.\ref{gammawindow} where $W$ becomes a constant independent of
energy and redshift. The anisotropies thus only depend on the
clustering properties of DM in the Milky Way (MW). We will take as a
reference model  the $C^l_{{\chi \rm gal}}$ from
\cite{SiegalGaskins:2008ge} where the anisotropies have been
calculated from a simulated MW substructure map. In
particular, we will use the spectrum shown in their Fig.5 for a
minimum clump mass of $10 M_\odot$. Compared with the cosmological
DM anisotropies this model gives a normalization of the angular
power spectrum a factor $\sim100$ larger (or more) depending on the
energy. More precisely, the shapes of the power spectra are similar for the two cases,
while the absolute normalization for the Galactic case is $l(l+1)C_l/2\pi\approx10$ at $l=100$
independent of energy, whereas for the extragalactic case
$l(l+1)C_l/2\pi\approx0.02$ at 10 GeV. Fig.\ref{fig:anieni} shows
the energy dependence of $l(l+1)C_l/2\pi$ at $l=100$ for the various
cases while Fig.\ref{fig:PSF} shows $l(l+1)C_l/2\pi$ for the DM EGB
case for 4 different energies.

\subsection{Model Sky Maps}
To produce a realistic simulation of data from Fermi we have to
simulate the expected sky pattern of every component. In the case of
Galactic foregrounds we employ the GALPROP package
\cite{Strong:1998pw,Moskalenko:1997gh} which produces as output the
sky maps  of the foregrounds as a function of energy. As already
described above the conventional diffuse model is used. It  is worth
noticing that energy dependent CR diffusion effects imply an energy dependent sky pattern as well.  In general, at low energies the diffusion length
of CRs is large and the resulting maps are thus smoother than the
higher energy counterparts where the diffusion length decreases.
Thus the relative importance of the Galactic foregrounds
anisotropies at small scales ($l\agt100$) increases with energy. Of
course, projection effects along the line of sight can sometimes
complicate the picture and give small scale anisotropies also at low
energies although this in practice happens only along the Galactic
plane where there is a contribution from structures many kpc away
from the solar position.

For the remaining components, i.e. the astrophysical EGB and the DM
EGB, we create template maps as gaussian realizations of the
theoretical angular power spectra $C^l$ derived above. The situation
is however complicated by the fact that the angular spectra are
energy dependent and, further, that there are non-zero
cross-correlation power spectra between different energy bands.
Energy dependent maps with the correct cross-correlation properties
among different energies need thus to be realized. In practice, we
choose $N_E=60$ logarithmically spaced
energy bands between 10 MeV and 10 TeV (i.e 10 bands per energy
decade). We then calculate the angular auto-correlation power
spectra for each band  and the cross-correlation power spectra
among all the possible pairs of energy bands according to equations
\ref{angspectra}-\ref{angcrossspectra}. The result is the model
power spectrum matrix $C_l^{ij}$ $i,j=1,N_E$ for the DM or
astrophysical EGB. We then generate the maps in harmonic space
($a^i_{lm}$, (see definitions in the appendix)) up to $l\approx 500$
sampling from the 60x60 matrix of the multi-gaussian distribution given
by the $C_l^{ij}$ for each $l$. In practice the energy behavior is
generally highly correlated so that only the first 4-5
principal components are required for the simulation of the 60 maps.
{The }  \verb+synfast+  {tool from HEALpix is then employed to combine the harmonic components
into the real space maps $f_{E_i}(\hat{\Omega})$}. These energy
dependent anisotropy maps are then weighted and normalized at each
energy according to the energy spectrum of the component itself
(astrophysical or DM EGB), giving the final template maps. These can
now be added to the model of Galactic foregrounds and to the model of
point sources to give a simulation of the gamma-ray sky where all
the components are entangled.

For the case of Galactic DM the procedure is simpler given that the
anisotropy pattern is independent of the energy and only one map
needs to be simulated. The same holds for the astrophysical EGB model with pure Poisson noise.

Convolution with the instrumental effects is described in the next section.

\section{The Fermi-LAT and simulated datasets}
In the following we describe the various instrumental effects
introduced in the Gamma-ray sky when observed through the eyes of
the Fermi-LAT telescope.

\subsection{Charged particles contamination}
Charged particles,  mainly protons, electrons and positrons, as well
as a smaller number of neutrons and Earth albedo photons, present a
major instrumental background to potential DM  signals, in
particular when considering the isotropic gamma-ray signal. These
background particles greatly dominate the flux of cosmic photons
incident on the  LAT, and a multivariate technique, employing
information from all LAT detector systems, is used to reduce the
remaining charged particle background to only a fraction of the
expected flux of extragalactic diffuse emission
\cite{Atwood:2009ez}.  The analysis of the extragalactic background
from Fermi \cite{Ackermann:prl} however indicates that, with the
standard data selection cuts, the residual contamination is
comparable to the photon background itself above a few GeV and
dominates at higher energies. Generally, this background is
expected to be basically isotropic so it can be easily taken into
account in the anisotropy analysis. Alternatively, stringent cuts
can be applied on the photon events decreasing the contamination but
also somewhat the exposure and thus the signal \cite{Ackermann:prl}.
For our simulation we will use the effective area for the
``diffuse'' event class as defined in \cite{Atwood:2009ez}. {Using
more stringent selection cuts to reduce the effects of CR
contamination reduces  the effective area by a
factor of $\sim$25\% \cite{Ackermann:prl}. The sensitivity curves
should change accordingly,  approximately by the small factor
$\sqrt{1.25}\sim1.1$.  }This estimate is reasonably accurate for a
sensitivity study as the present one. We do not attempt a detailed
simulation of the CR background itself, postponing a more careful
investigation of the issue to the data analysis work.

\subsection{Point Spread Function and Exposure}

The effects of the Fermi-LAT detector can be summarized in terms of
the IRFs  which include the point
spread function (PSF) and the energy dependent exposure maps. For
this analysis we use the PSF and exposure as available from the
Fermi Science tools p6v3 \footnote{See
{http://fermi.gsfc.nasa.gov/ssc/} }. We neglect the effect of energy
dispersion (which is about 10 \%) and assume that the photon energy
is perfectly reconstructed. In this respect, for this analysis we
will divide the simulated datasets in wide energy bands so that this
instrumental effect is negligible for our purposes.  To obtain
simulated maps we use the energy dependent model input maps and then
convolve them with the Fermi PSF and simulated exposure of 5 years.
{More specifically, the exposure was generated from 1 yr of pointing history 
simulated with } \verb=gtobssim=  {and then rescaled to 5 yrs. The convolution with the PSF and the exposure 
was then performed using the \emph{GaDGET} \cite{Ackermann:2009zz} tool (see also Abdo et al. (2009b)). } 
The resulting maps have a normalization in units of counts per
steradian which is further re-normalized to counts per pixel ($\sim
10^{-5}$sr for a Nside=256 HEALpix pixelization). Finally Poisson
random noise is added to simulate the effect of finite statistics.

The Fermi PSF is energy dependent and improves at high energies. It
corresponds to an angular resolution of about $3^\circ$ at 100 MeV,
$0.6^\circ$ at 1 GeV and $0.1^\circ$ above 10 GeV (see
\cite{Atwood:2009ez} for more details). The PSF thus heavily affects
the anisotropies especially below 1 GeV, suppressing the medium-high
multipoles in the angular power spectrum. The sensitivity to these
multipoles is not completely lost however. In principle, from
the knowledge of the PSF it is possible to recover the true power
spectrum even at higher multipoles although the error grows with
larger attenuation (it grows exponentially with multipole $l$: the suppression factor is given by  
$W_l= exp(-l^2\sigma_b^2/2)$, for a gaussian PSF with variance $\sigma_b$ ).

\begin{figure}
\vspace{-1.0pc}
\begin{center}
\begin{tabular}{c}
\vspace{-0.5pc}\includegraphics[width=0.99\columnwidth,angle=0]{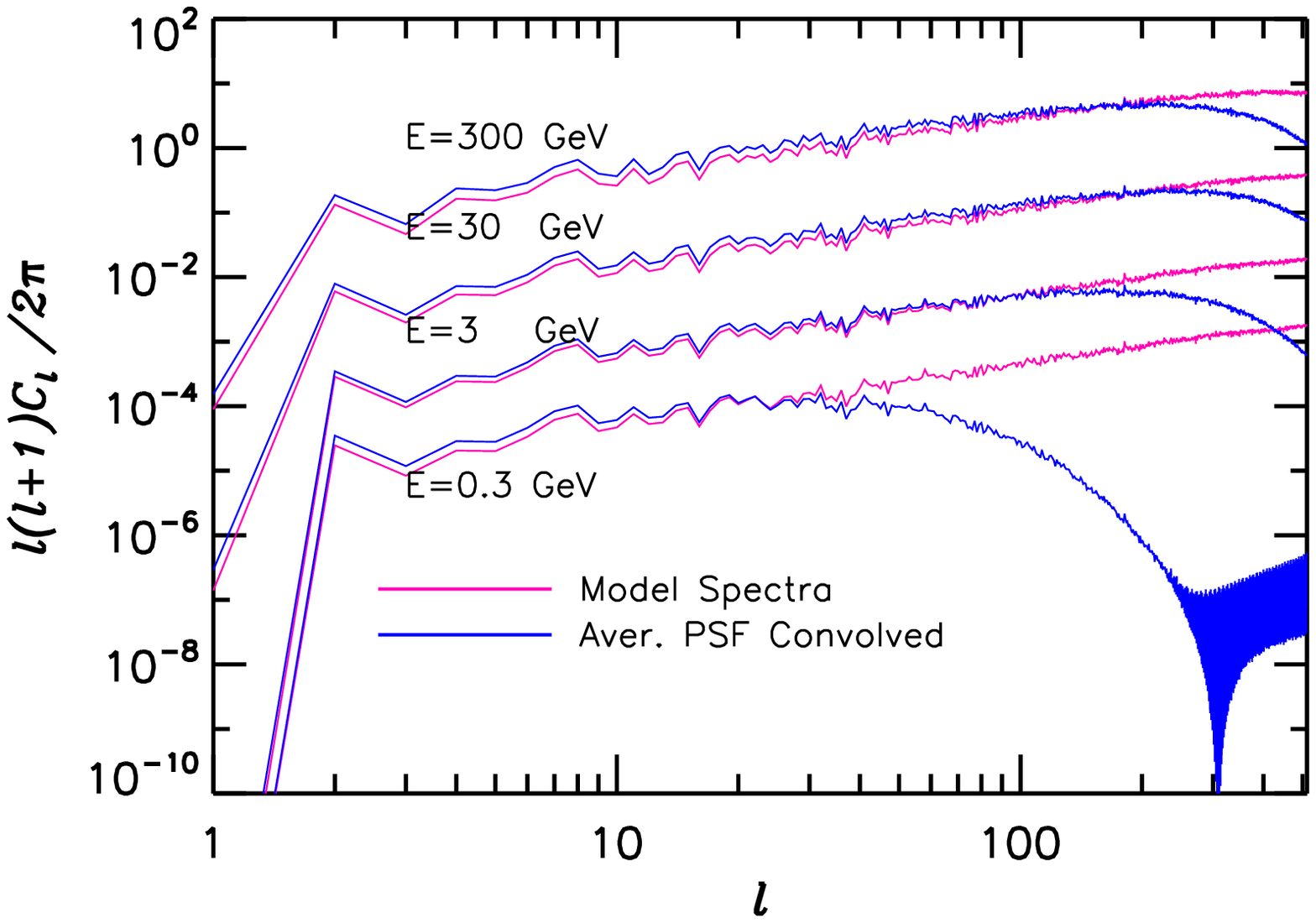} \\
\includegraphics[width=0.99\columnwidth,angle=0]{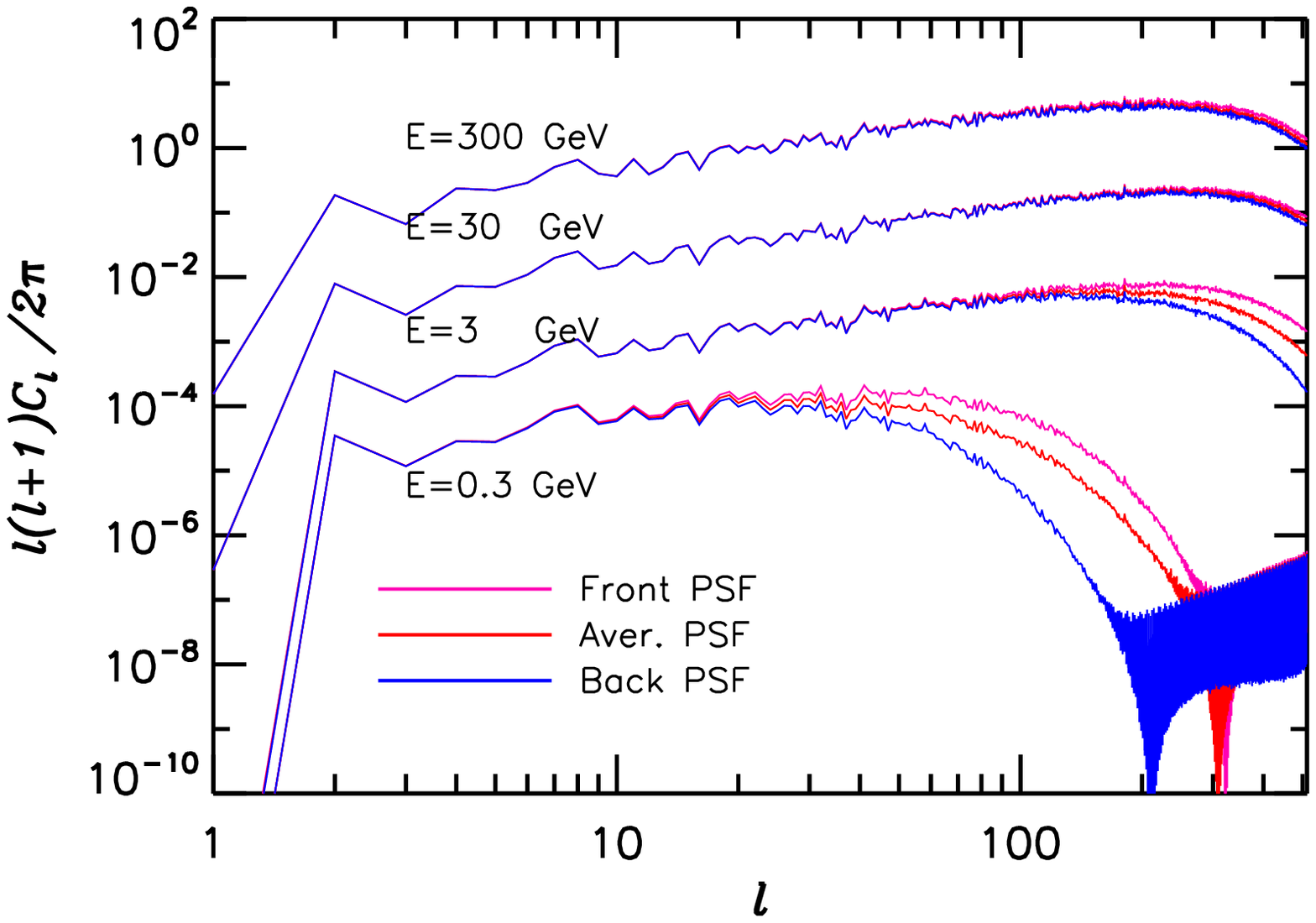}
\end{tabular}
\end{center}
\vspace{-1.5pc} \caption{Power spectra at various energies of the
model of DM EGB emission (for a $m_\chi$=1 TeV WIMP annihilating
into $b\bar{b}$) and their corresponding PSF convolved
counterparts (from top to bottom at $E=300$ GeV, $E=30$ GeV, $E=3$
GeV, $E=300$ MeV). The PSF suppresses anisotropies at $l>100$ for
E\alt 1GeV. The rise of the power spectrum at high multipoles in
the $E=300$ MeV case is a numerical HEALpix artifact.  The lower
panel shows the power spectra resulting from the different PSFs for
events converting in the front or the back of the detector and for
the average PSF. The dipole has not been included in the simulated maps,
hence the sharp rise at $l=2$. \label{fig:PSF}}
\end{figure}

The effect of the PSF can be seen in  Fig.\ref{fig:PSF} where our
model power spectra for the DM EGB component at various energies are
compared with the corresponding PSF convolved spectra. The latter
are obtained by convolving the model DM EGB sky maps (simulated from
the angular spectra with the procedure described in the previous
section) with the PSF  and extracting the power spectra from the
resulting maps. As expected, the PSF heavily suppresses anisotropies
at $l>100$ for E\alt 1GeV. Notice also that there is a certain
suppression at $l\agt 200$ even for $E=300$ GeV, which is  slightly more than 
the factor $W_l= exp(-l^2\sigma_b^2/2)$ expected at this energy for a gaussian PSF with width $0.1^\circ$. 
This extra effect is likely related to the
non-gaussian form of the PSF which even at very high energies
($>100$ GeV) exhibits a tail towards large angular spreads. The
figure also shows the difference in PSF for events converting in the
front or the back of the detector where the reconstruction
performances are different. It can be seen that the PSF is better
for events converting in the front, giving a sensitivity to slightly
higher multipoles especially at low energies. Above a few GeV,
instead, the difference is less important. We will use the average
PSF for our analysis.

The exposure maps are very uniform over an averaging period of 1
year (see again \cite{Atwood:2009ez}) so that the convolution with
the exposure only adds power to the very lowest multipoles
$l\alt10$. This can seen in fig.\ref{fig:3} where we plot the
angular power spectrum of the exposure which, indeed, falls rapidly
at $l=10$. The exposure pattern is also energy dependent, although
only weakly. The integrated exposure, however, rises steeply starting
from about 100 MeV and then flattens above about 1 GeV.

\begin{figure*}
\vspace{-1.0pc}
\begin{center}
\begin{tabular}{cc}
\vspace{-0.2pc}\includegraphics[width=0.56\columnwidth,angle=90]{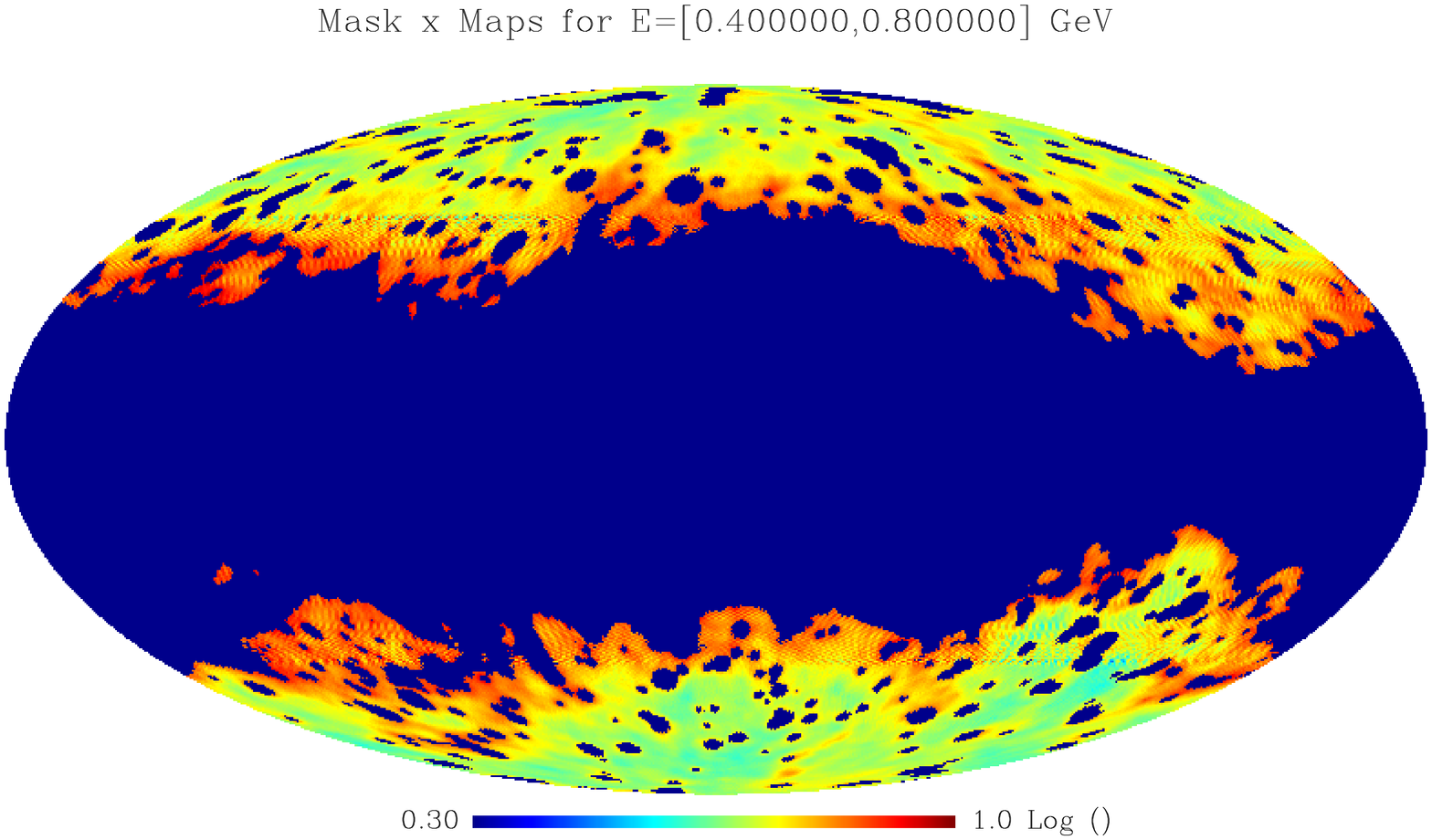}&
\vspace{-0.2pc}\includegraphics[width=0.56\columnwidth,angle=90]{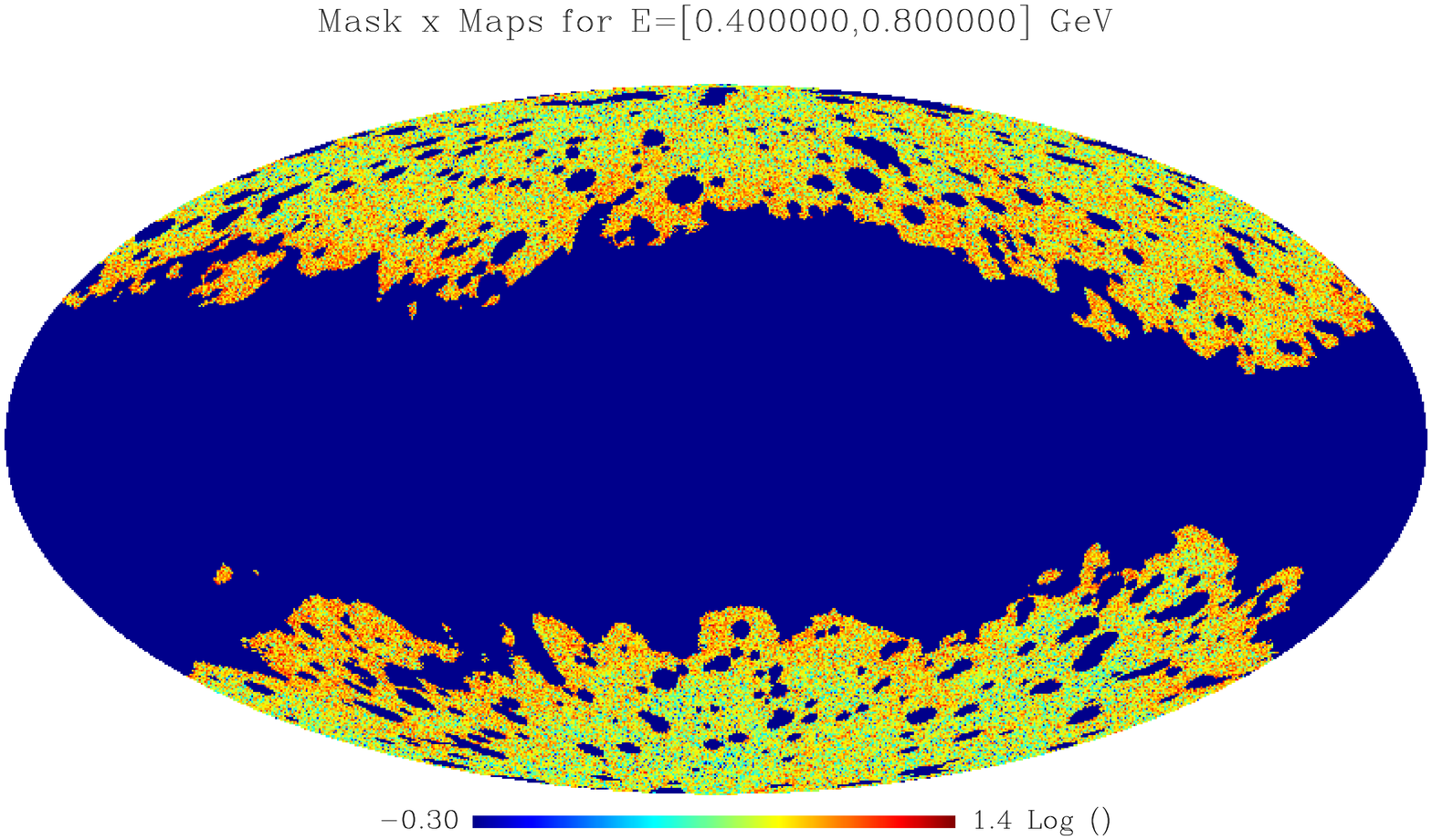}  \\
\vspace{-0.2pc}\includegraphics[width=0.56\columnwidth,angle=90]{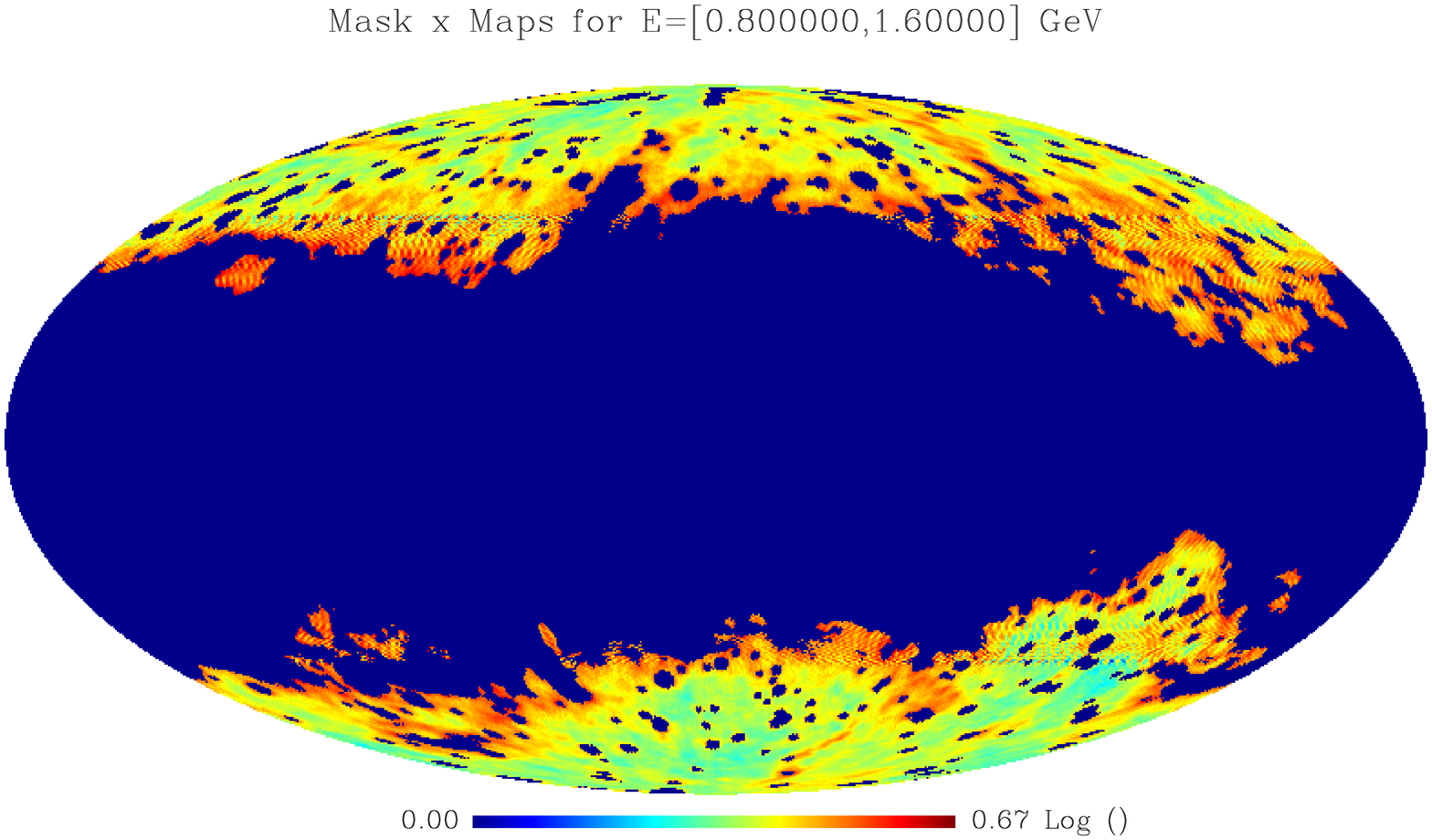}&
\vspace{-0.2pc}\includegraphics[width=0.56\columnwidth,angle=90]{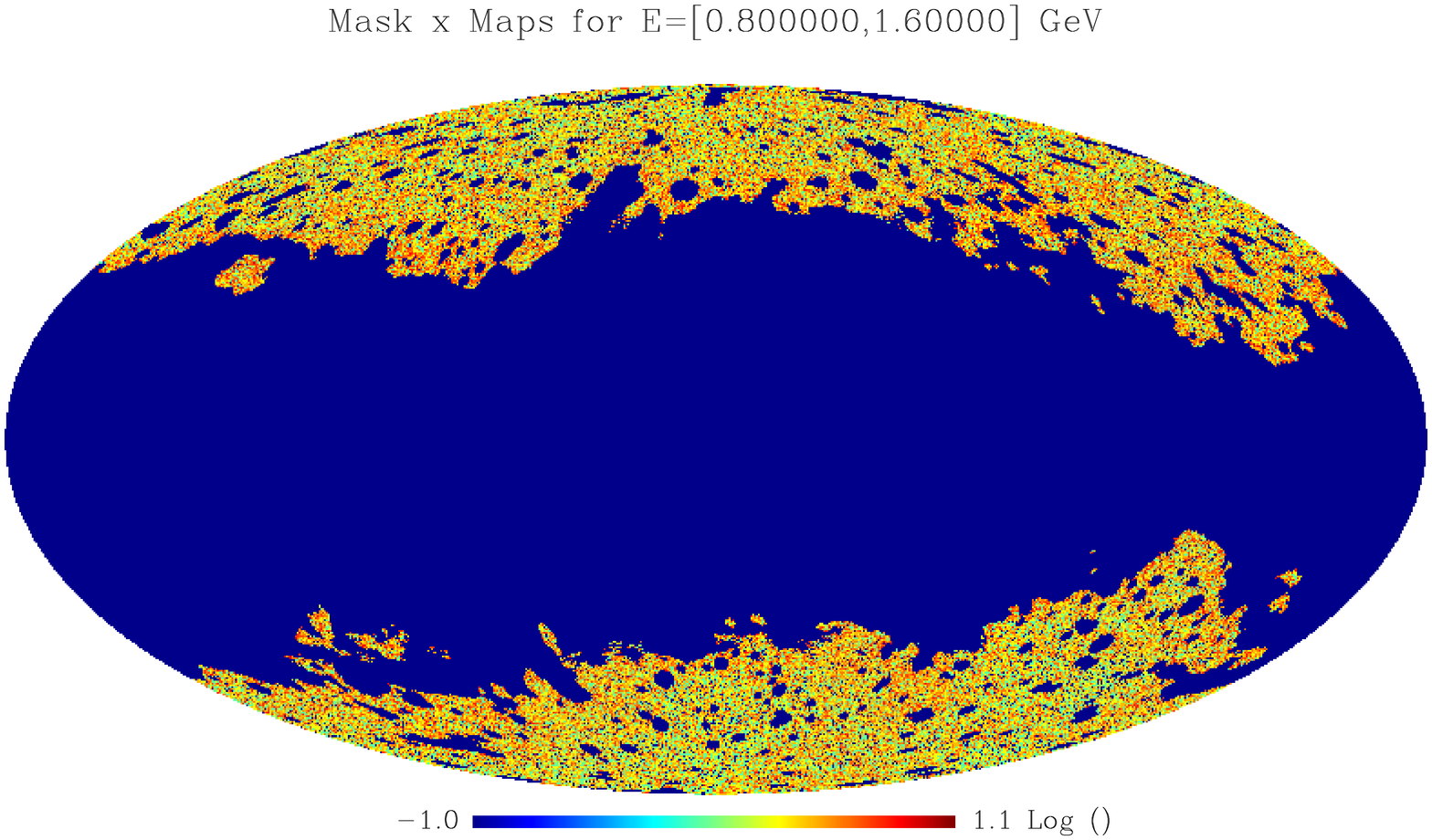}  \\
\vspace{-0.2pc}\includegraphics[width=0.56\columnwidth,angle=90]{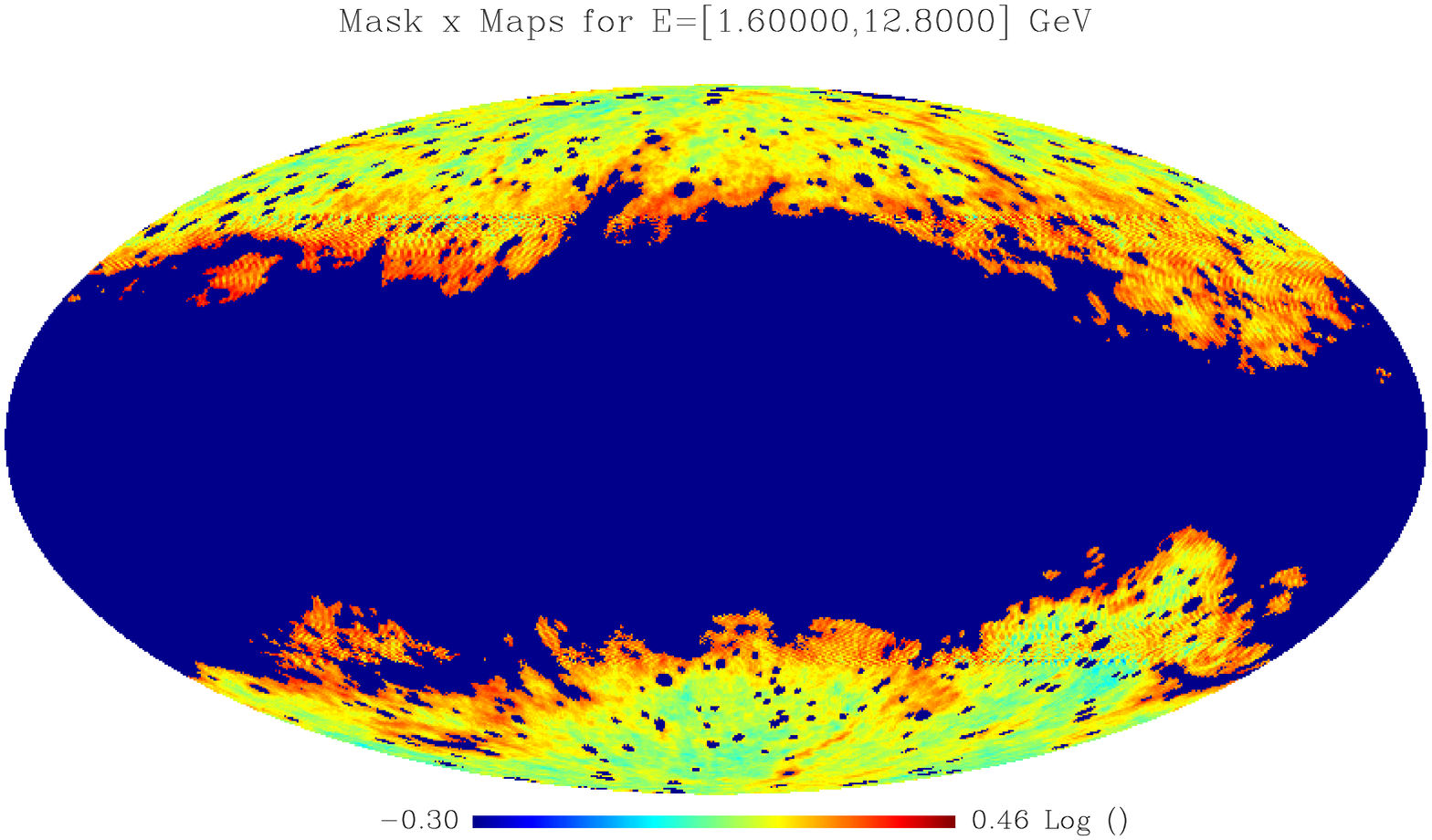}&
\vspace{-0.2pc}\includegraphics[width=0.56\columnwidth,angle=90]{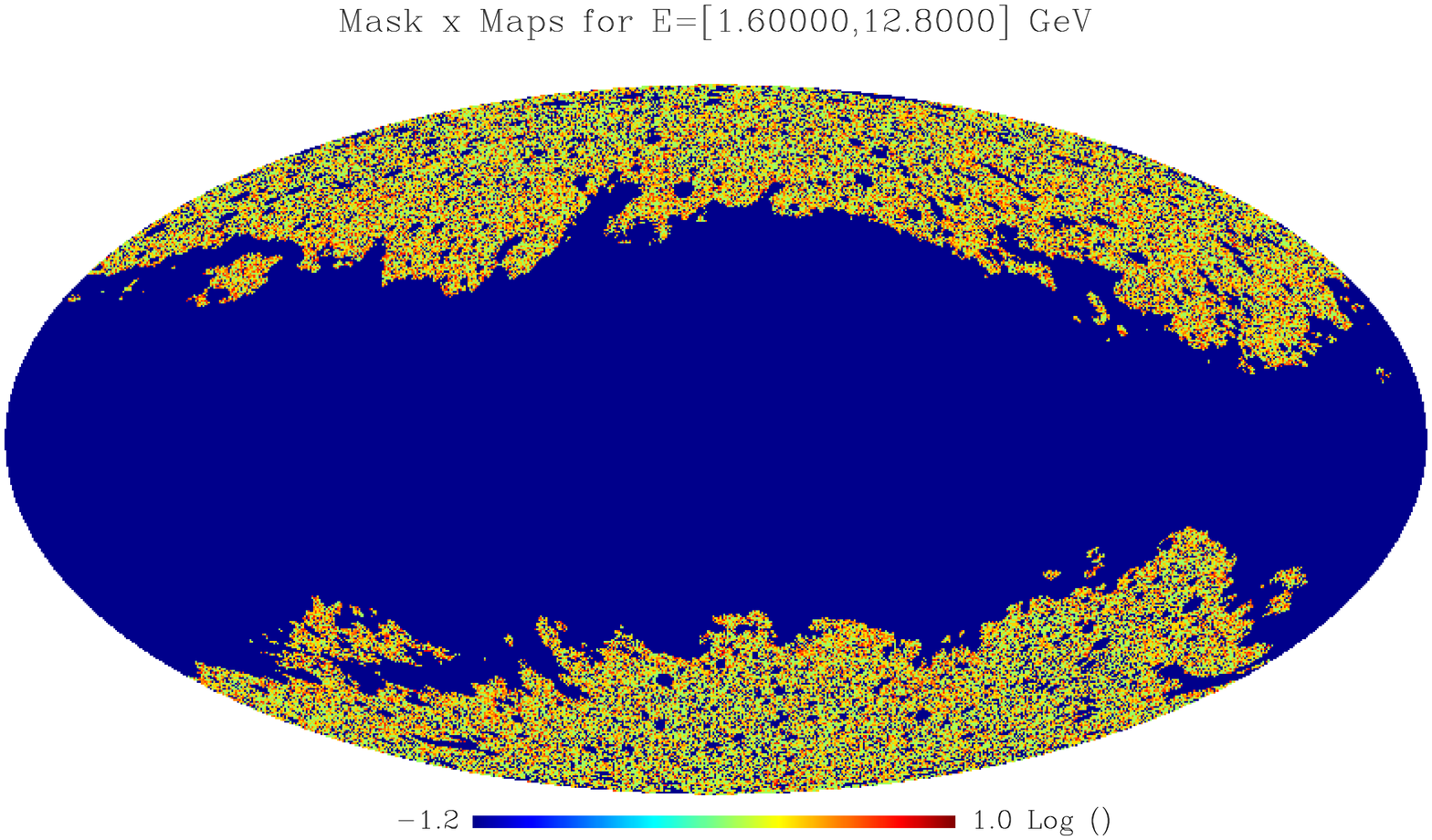}  \\
\vspace{-0.2pc}\includegraphics[width=0.56\columnwidth,angle=90]{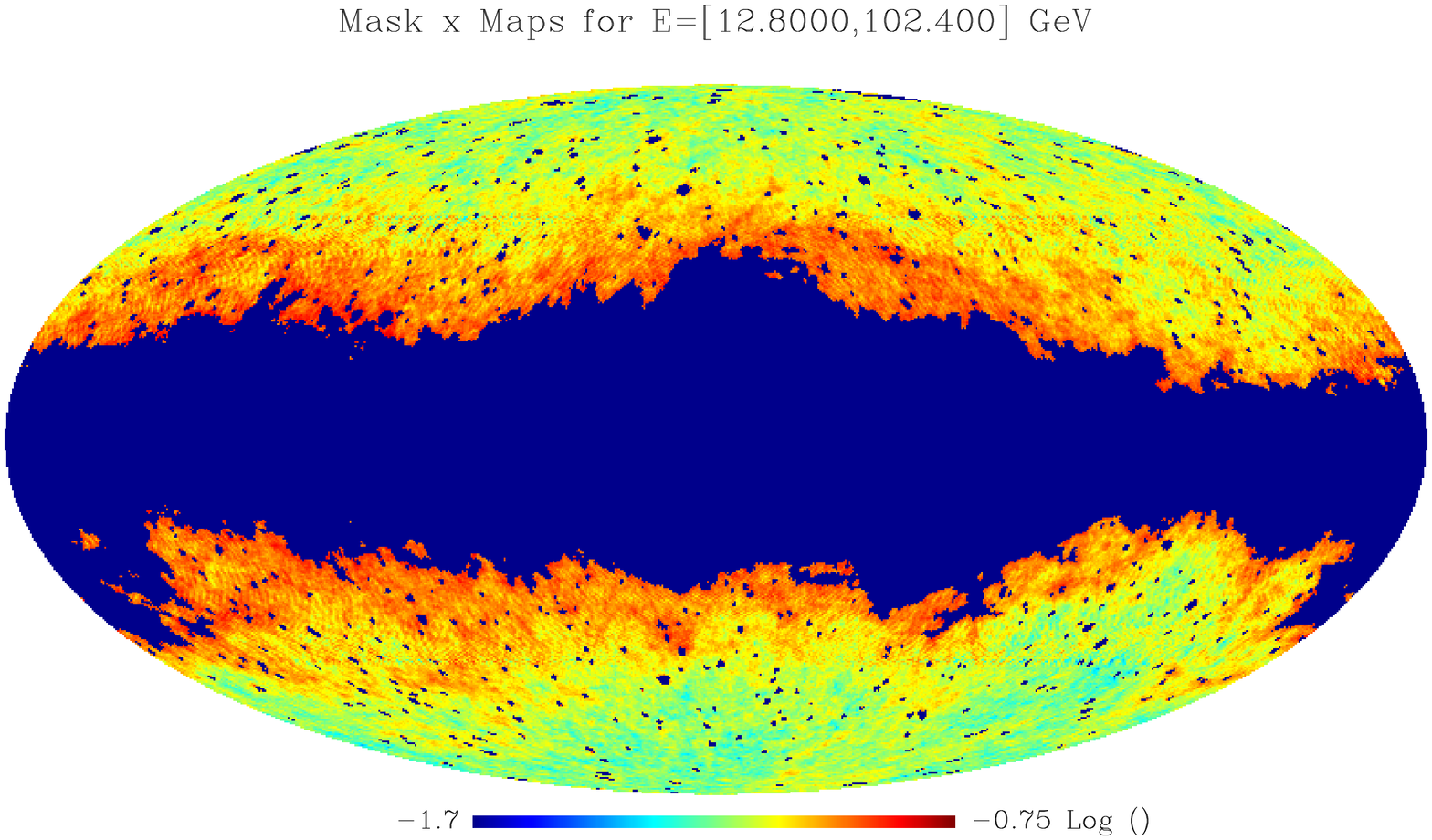}&
\vspace{-0.2pc}\includegraphics[width=0.56\columnwidth,angle=90]{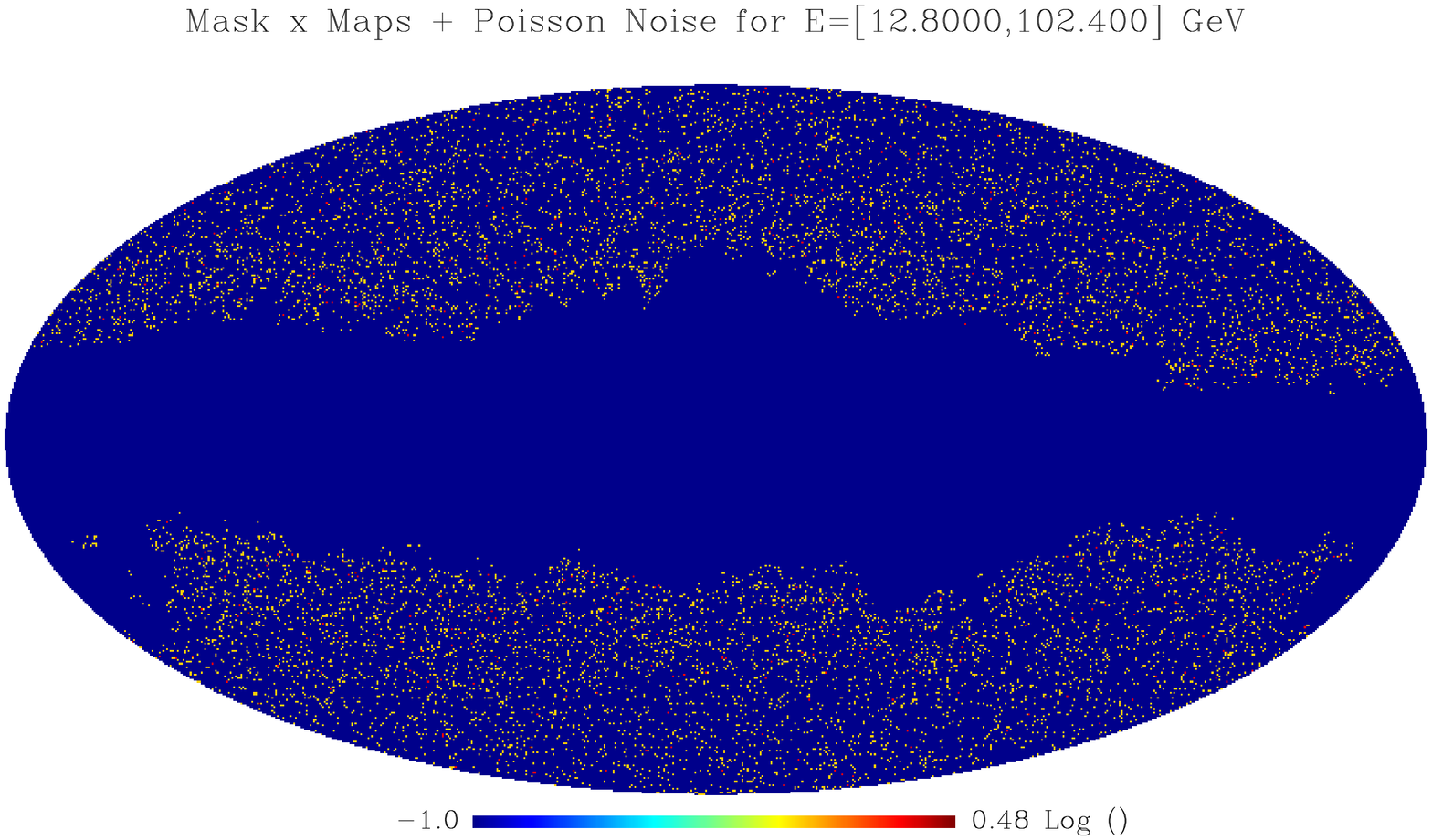}  \\
\vspace{-0.2pc}\includegraphics[width=0.56\columnwidth,angle=90]{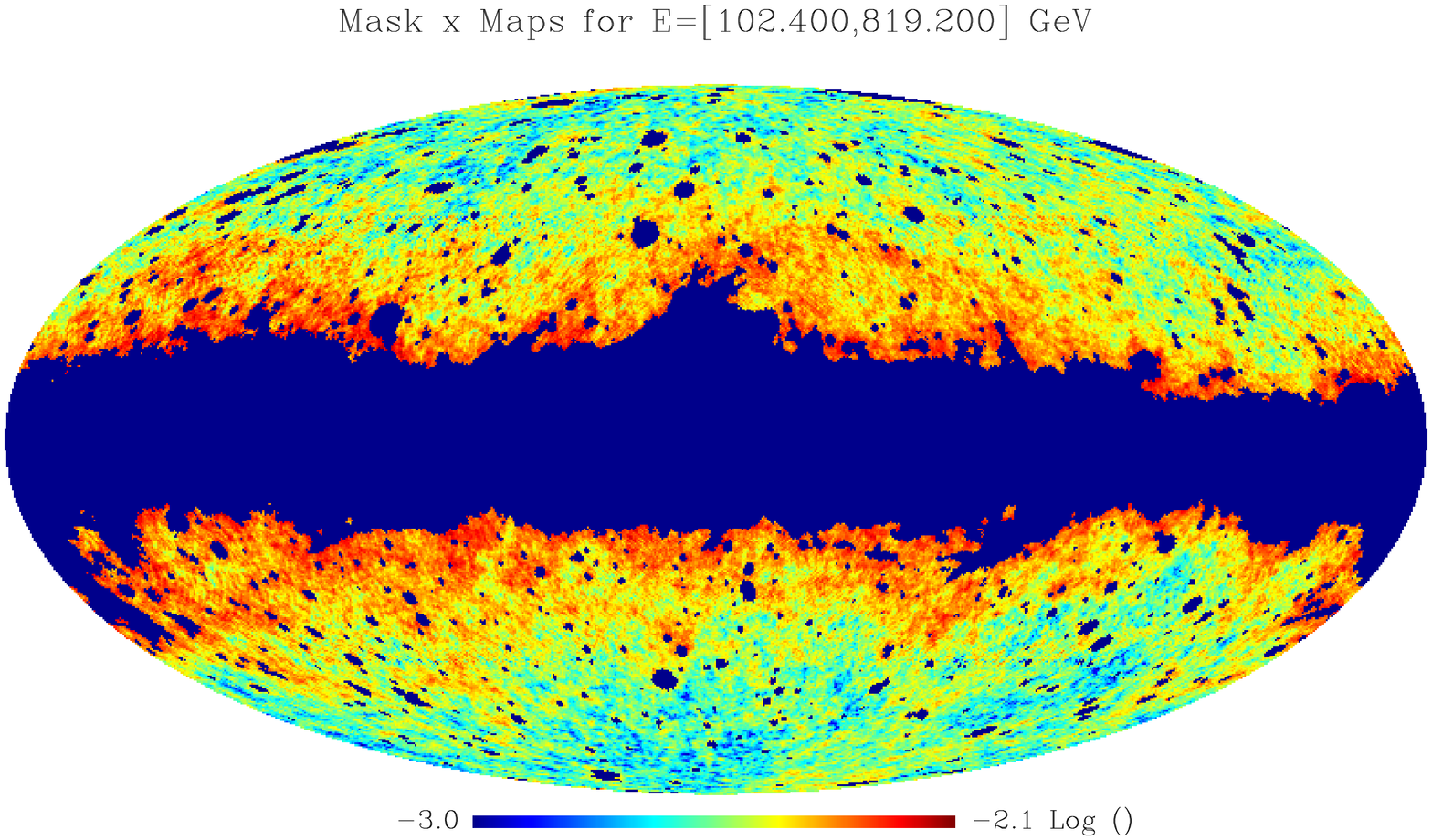}&
\vspace{-0.2pc}\includegraphics[width=0.56\columnwidth,angle=90]{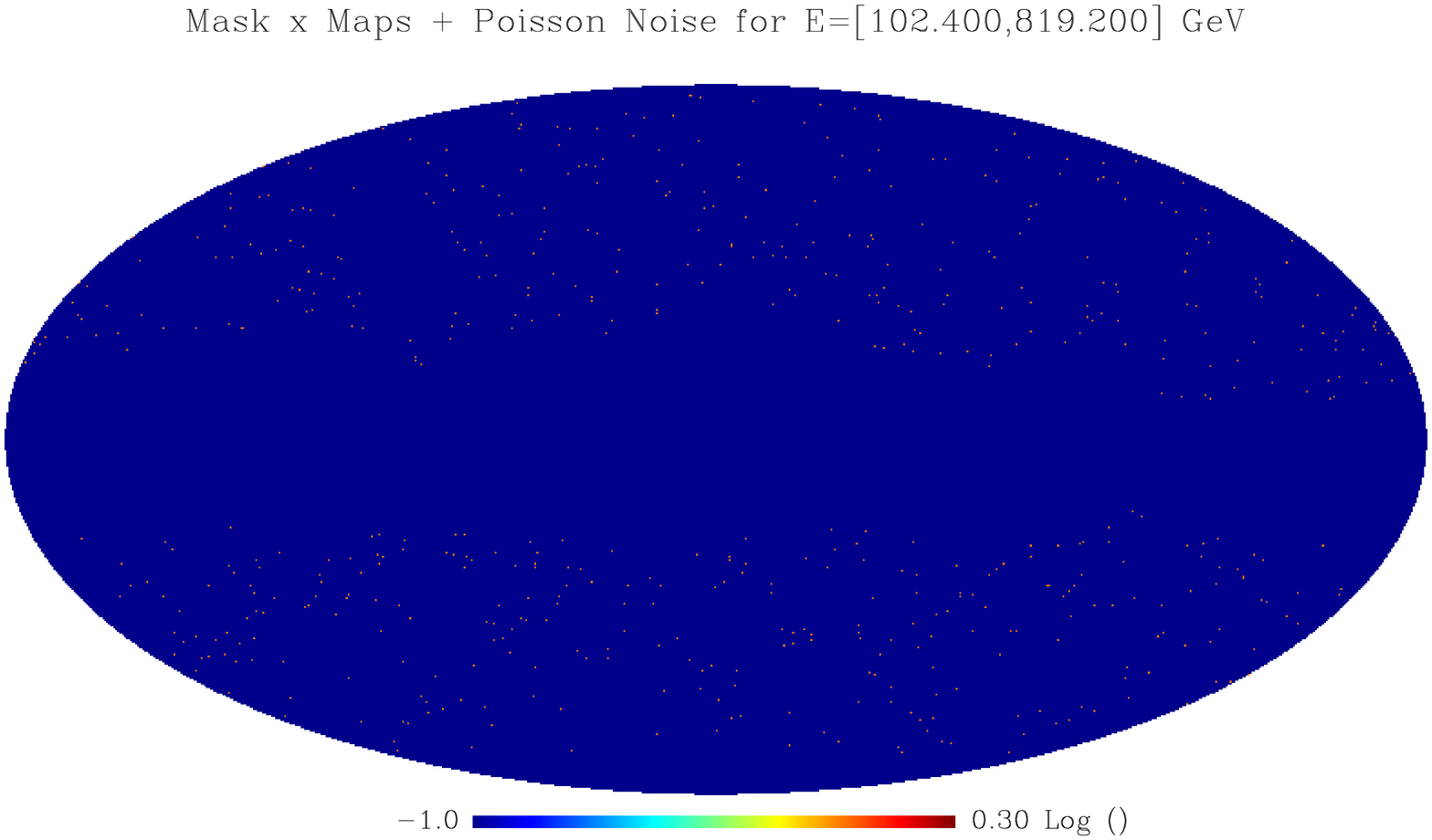}  \\
\end{tabular}
\vspace{-0.5pc}
\end{center}
\hspace{-1.5pc} \caption{(Left) 5 year averaged counts map (in HEALpix
nside=256 format) of our reference no-DM model, which
contains astrophysical EGB, Galactic foreground and resolved point
sources. An energy dependent mask is applied to suppress Galactic
foregrounds and point sources (see the text). The simulation has
been divided into five energy bands (in the energy ranges indicated
in the titles) and only the region outside the masks is shown. The
EGB energy spectrum is harder than the Galactic diffuse one, so at
high energies there is more sky area available for the analysis.
(Right) A random realization of the expected counts. \label{fig:8}}
\end{figure*}

\begin{figure*}
\hspace{-8.5pc}
\hspace{-0.5pc}\includegraphics[width=2.40\columnwidth,angle=0]{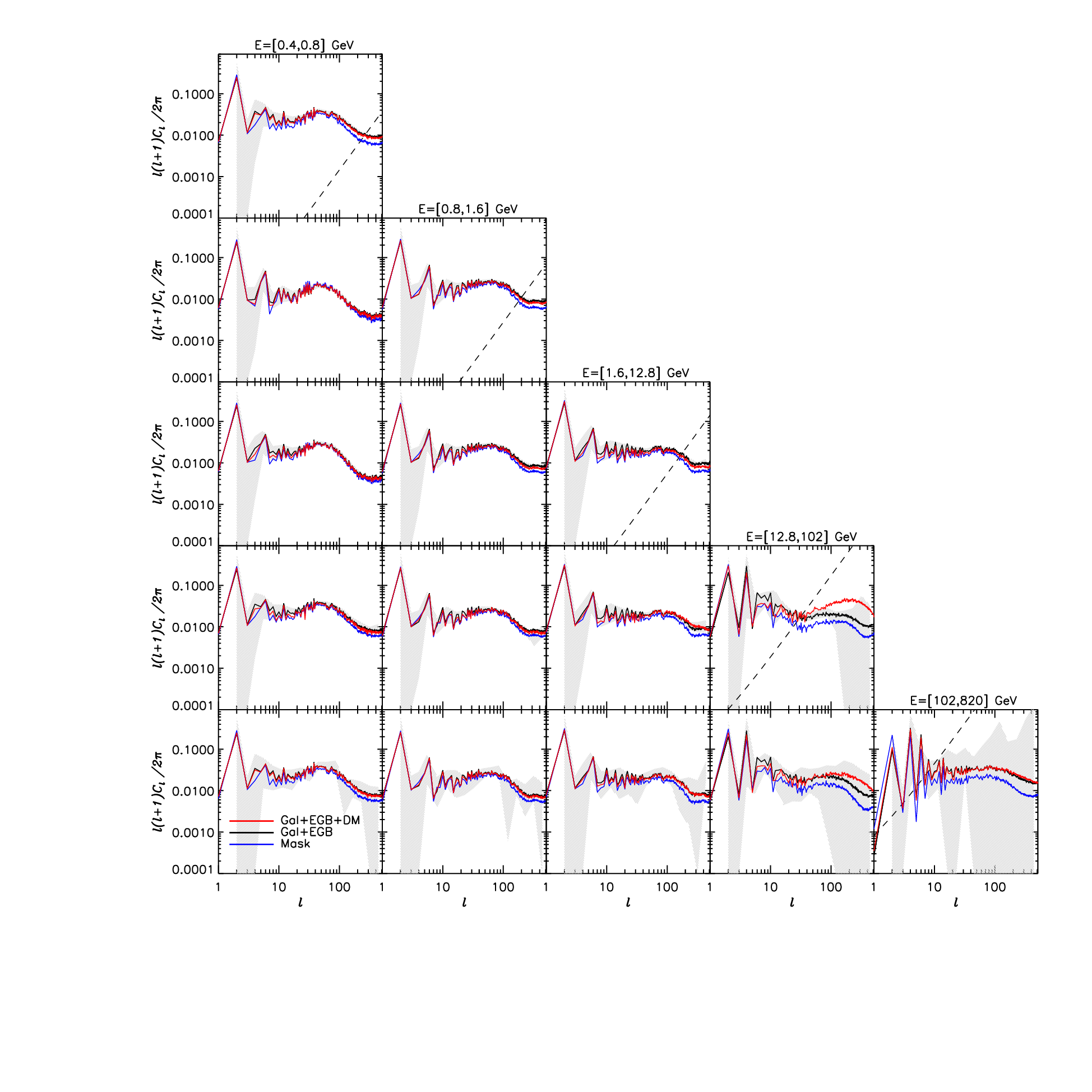}
\vspace{-0.5pc}
\vspace{-7.5pc} \caption{Power spectra matrix for  the 5 average
counts maps (times the masks) of the previous figure. Before computing the spectra, the maps have
been (re)normalized so that to represent fluctuation maps. Auto-correlations are
in the diagonal and cross-correlations are off-diagonals. The dashed lines in the auto-correlation plots
indicate the level of shot noise $\propto 1/N_{\rm counts}$. The
noise in different energy bands, in contrast, is uncorrelated so that no
shot noise is present in the cross-correlations. The blue lines
represent the power spectrum matrix of the masks alone. It can be
seen, that the masks give the ``bulk'' level of
anisotropies. The anisotropies of the masked no-DM reference model
then add over the mask level (black curves).  At higher energies
the angular resolution of Fermi-LAT increases and
the anisotropy excess with respect to the masks is better visible
(in the lower right part of the plot) at medium-high multipoles. To
be at a significant level, the difference has to exceed
the errors bars. Statistical errors
are represented by the gray shaded area for a confidence level of two sigma. 
Finally, the red curves represent the angular
power spectra when a DM EGB component is added on top of the
previous contributions. For illustration $m_\chi \approx 200$ GeV is
chosen and the DM flux is normalized to the same level as the
astrophysical EGB at 10 GeV. The bump visible above the statistical error
level in the $10$ GeV-$100$ GeV autocorrelation spectrum (and the absence
of it in the other energy bins) would make this model
detectable. \label{fig:9}}
\end{figure*}

\section{Results of the Simulation}

\subsection{Simulated Sky-Maps}

The maps produced as described in the previous sections are summed up
 to build a simulation of the observed gamma-ray sky. A
reference no-DM model is derived adding Galactic foregrounds,
astrophysical EGB and point sources. The astrophysical EGB is
normalized to the level of the observed IGRB. The normalization of
the Galactic foregrounds is the one given as output by GALPROP.
Point sources are normalized to their observed flux. The overall
statistics is normalized to 5 years of data taking by
Fermi. The maps we obtain are \emph{average counts maps}, but a
realization of \emph{effective counts maps} can be easily obtained
drawing poisson counts in each pixel from the average counts
maps. Although we use 60 energy bands to produce the simulated maps,
it is convenient to rebin them into fewer maps, in order to have a
reasonable amount of statistics in each band. We thus divide the
dataset in 5 energy bands from 400 MeV to 800 GeV. Finally, a
crucial step is the construction of the masks which will suppress
the Galactic foregrounds and the resolved point sources. For each
energy band  the mask are defined as the region where the
foregrounds do not exceed twice the EGB (200\%), and the point
sources emission does not exceed more than 20\% of the EGB itself.
Simulated sky maps (outside the masked regions) for both the average
counts and one particular realization of the counts are shown in
Fig.\ref{fig:8}. Indeed, outside the masks the remaining signal is
quite isotropic, indicating that the IGRB makes most of the signal,
but still a certain residual foreground contamination is apparent,
especially near the edges of the mask. Clearly this is expected
given the loose cut employed to build the masks (only 200\% of the
EGB). A tighter cut on the foregrounds can be applied but then the
available area for the analysis would shrink to close to zero since
the Galactic foregrounds never go below 20-30\% of the EGB even at
the Galactic poles. For the present forecast analysis a substantial
foreground contamination is still acceptable since the foreground
model is known in advance and what is required for a significant
detection is that the EGB anisotropy exceeds the anisotropy of the
residual Galactic foregrounds. In general, for a real data analysis
some kind of foreground cleaning must be performed to obtain useful
results. We postpone a detailed treatment of this issue to the analysis of the
real data.

On the other hand, the cut of 20\% of the EGB for point sources is
rather effective and removes most of the gamma-ray events from
resolved point sources, except perhaps few events around the most
powerful sources where, statistically, some events very far from the
source position are expected due to the tails of the PSF. From
Fig.\ref{fig:8} it can also be seen that the fraction of sky where
the isotropic component dominates increases with energy. This is expected
considering the fact that the IGRB has a spectrum harder than the
Galactic emission (see also Fig.\ref{fig:diffuse}). Also it is clear
that the area masked for the point sources substantially decreases
at high energy due to the improvement of the PSF. This is true for
all but the highest  energy band, where the fraction of sky covered
by the point sources increases again. This can be understood looking
at the energy spectrum of the resolved sources in
Fig.\ref{fig:diffuse}. This becomes very hard for $E\agt 10$ GeV so
that, even if the PSF at this energies is very narrow, its tail
still contains a flux which is a significant fraction of the IGRB.

\subsection{Angular Power Spectra}

In Fig.\ref{fig:9} we plot the various auto-correlation and
cross-correlation spectra in a ``triangle plot'', where the diagonal
represents the auto-correlations and the the off-diagonals the
cross-correlations. The blue lines represent the spectra of the
masks only, which thus provide the zero-level anisotropy. The black
line represents the average spectrum of the reference no-DM model
convolved with masks, i.e. the maps in the left column in
Fig.\ref{fig:8}. As explained above, outside the mask the reference
model is to a large extent given by the astrophysical EGB, although a
substantial residual contamination from the Galactic foregrounds is
still present. The residual point source contamination, on the other hand,
is negligible. The grey shaded area is the two sigma error
region in the power spectrum accounting for the effect of the finite
number of events. A logarithmic binning of the multipoles in 18 bands is
used. The level of shot noise is shown with a dashed line. It can be
seen that the shot noise increases in the high energy bands where
the statistics decrease, as expected. Finally, the red curves
represent the angular power spectra if a EGB DM component is added
on top of the previous contributions. For illustration  a WIMP with $m_\chi
\approx 200$ GeV is chosen and the DM flux is normalized to the same
level of the EGB at 10 GeV. It can be seen, indeed, that in the 4th
energy band (approximately 10-100 GeV), where the DM emission peaks,
the anisotropy has a noticeable bump compared to the EGB anisotropy
and provides a clear signature for this model. The DM anisotropy
signal disappears in the 5th band (100-800 GeV) where the DM
contribution drops. This ``energy modulation in
the anisotropy spectrum'' is also discussed in  
\cite{Hensley:2009gh}).

It has to be noted that the spectra look quite different from the
all-sky spectra of the astrophysical and DM EGB shown in
Fig.\ref{fig:3}, both in the shape and normalization. This is of
course due to the effects of the residual foregrounds  and
especially of the mask which distorts the original spectrum. The
mask effects can in principle be removed with specifically designed
algorithms like the MASTER algorithm \cite{Hivon:2001jp}. This
comes, however, at the price of further errors in the final spectrum
which need to be addressed with Montecarlo simulations (a package
implementing the full procedure is described for example in
\cite{Lewis:2008wr} ). For the present purpose, it suffices to use
the masked sky power spectrum without attempting to deconvolve it
from the mask itself. This is fine as long as all the power spectra
are calculated on the same sky region in order to allow a consistent
treatment.

\begin{figure}
\vspace{-0.0pc}
\begin{center}
\begin{tabular}{c}
\hspace{0.0pc}\includegraphics[width=0.90\columnwidth,angle=0]{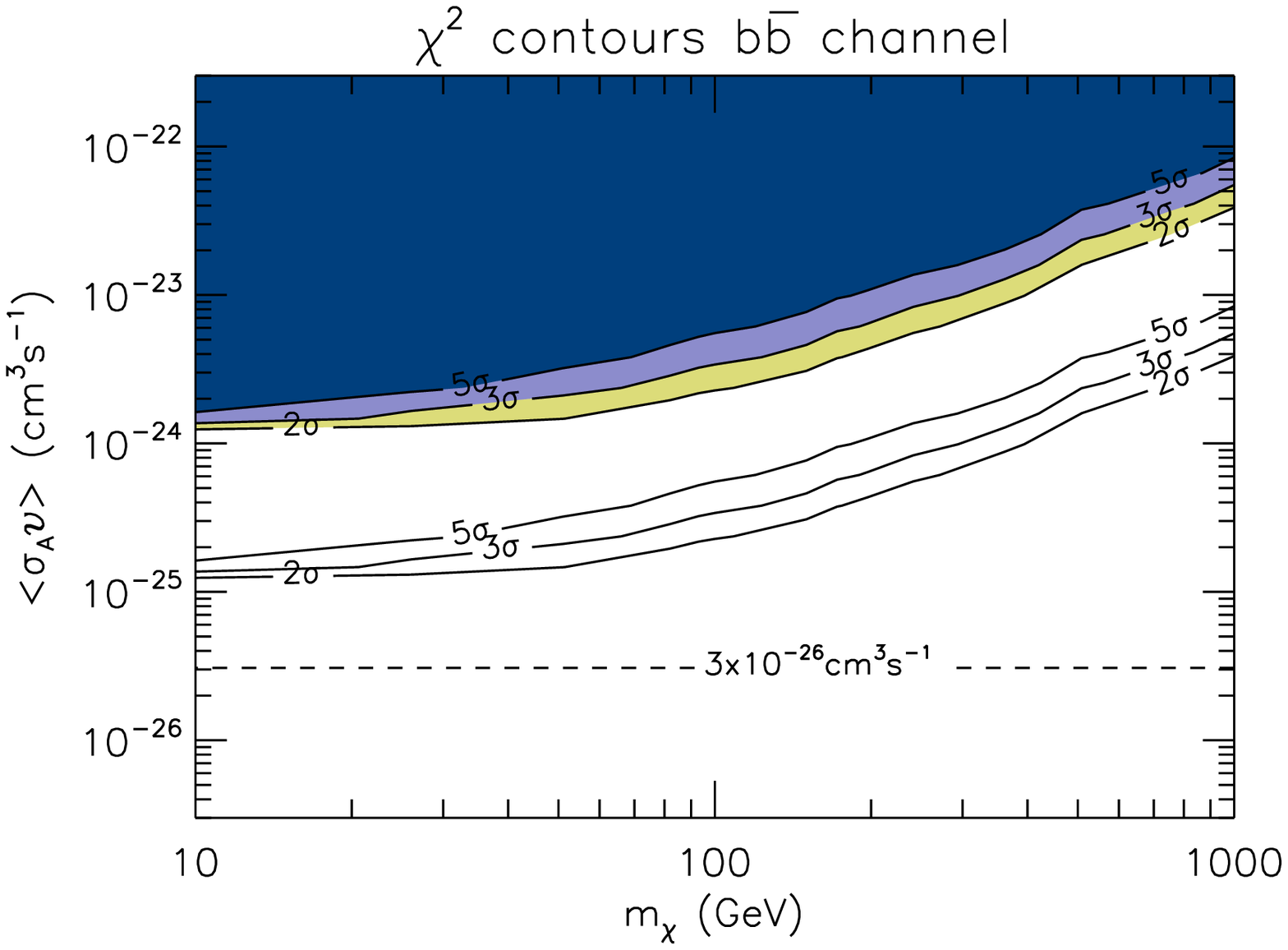}  \\
\vspace{-0.0pc}\includegraphics[width=0.90\columnwidth,angle=0]{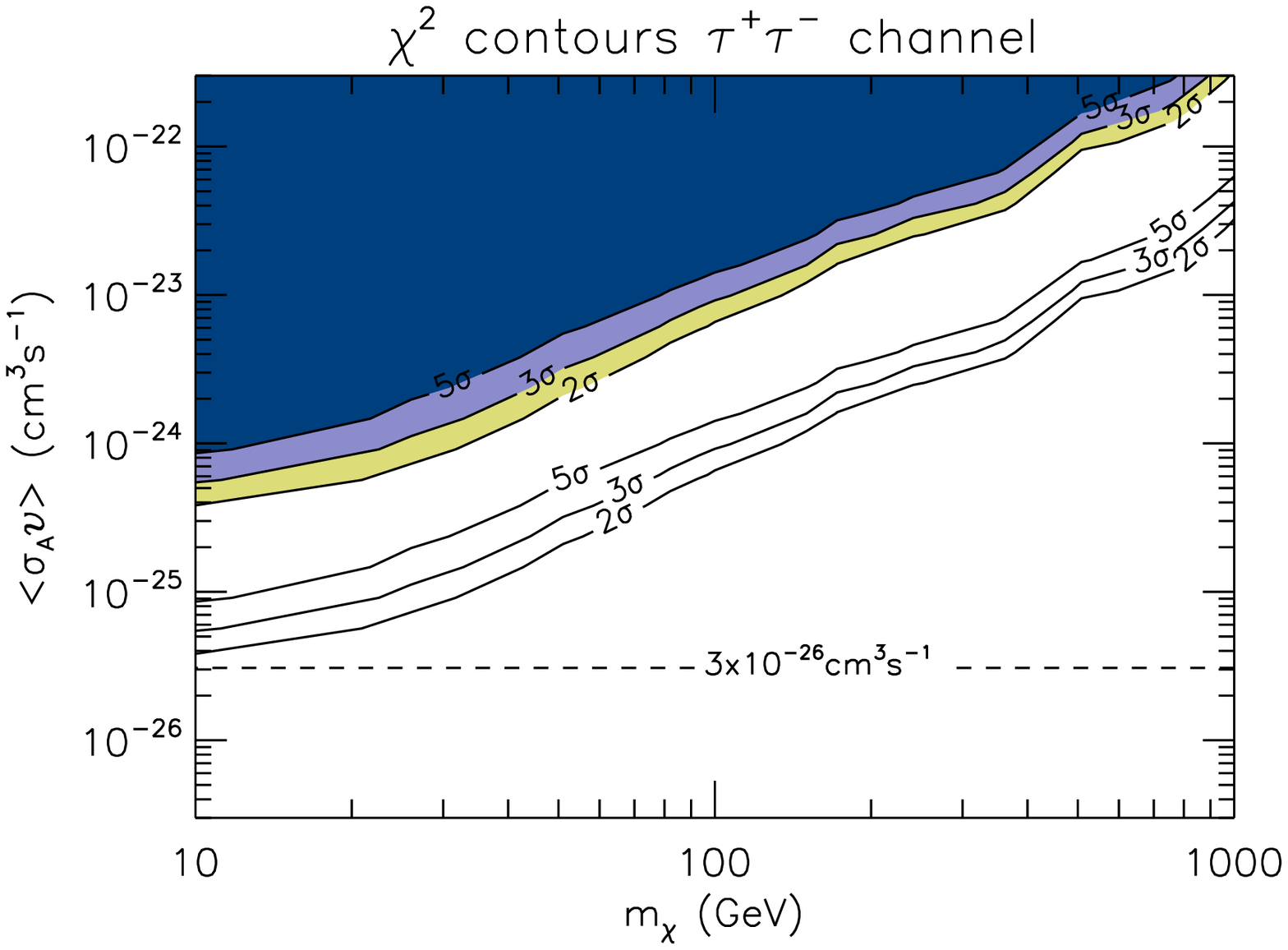} \\
\hspace{0.0pc}\includegraphics[width=0.90\columnwidth,angle=0]{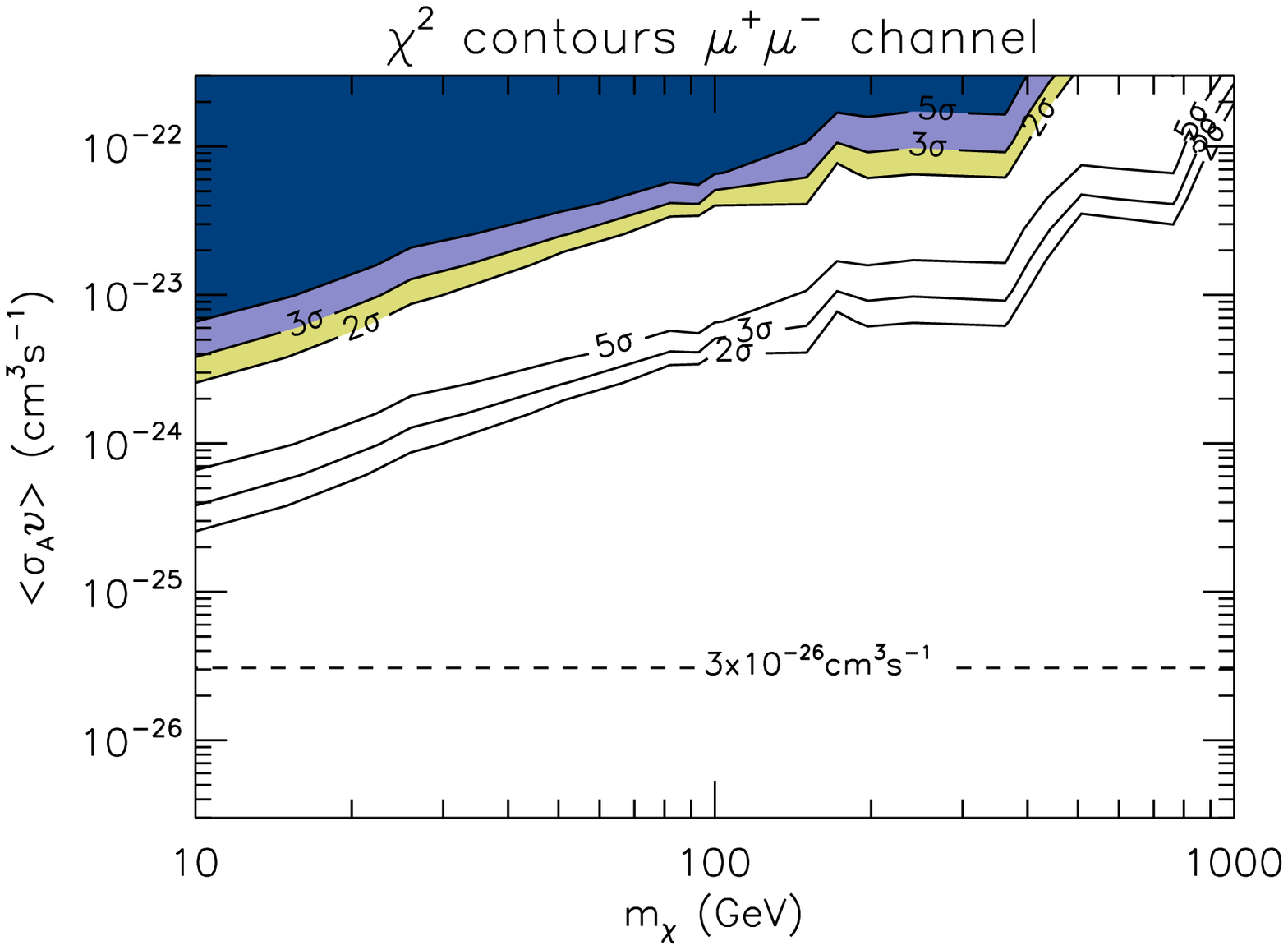}  \\
\end{tabular}
\vspace{-0.5pc}
\end{center}
\hspace{-1.5pc} \caption{Exclusion plots in the $m_\chi$-$\left<
\sigma_{\rm{A}}v\right>$ plane for various annihilation channels:
$b\bar{b}$, $\tau^+\tau^-$, $\mu^+\mu^-$. Each
panel shows the sensitivity curves at the $2\sigma$, $3\sigma$ and
$5\sigma$ levels. Two different normalizations for the flux of the
cosmological DM signal have been employed using an updated version
of the (Ullio et al. 2002) model: NFW Halo profile with
substructures (lower curves) and no-substructures (higher curves).
The curves are the sensitivities considering the auto-correlation
spectra information alone. \label{fig:limitsa}}
\end{figure}

\section{DM sensitivity }\label{sec_sense}

It is in principle straightforward
to calculate the sensitivity to DM requiring that the DM anisotropy 
does not exceed the anisotropies of the reference no-DM
model. Let us define
$C_{EGBl'}^{ij}$ as the binned multipoles power auto- and
cross-correlation spectra of the reference model \footnote{We label
it for simplicity EGB although, besides the true
astrophysical EGB, there are some residual contaminations from
Galactic foregrounds and point sources (see previous section).}. The
indices $i,j$ run from 1 to 5 and indicate the energy bands and the
index $l'$ indicates the multipoles bins. Let us also define $\Delta
C_{EGBl'}^{ij,  mn}$ as the covariance matrix of the
$C_{EGBl'}^{ij}$, i.e. $\Delta C_{EGBl'}^{ij,  mn}$ is the
covariance between $C_{EGBl'}^{ij}$ and $C_{EGBl'}^{mn}$.  We divide
the multipole range $2-512$ into 18 logarithmically spaced multipole
bands.  Adding to the previous signal, the contribution for a given
DM model, specified through the annihilation channel, the velocity
averaged annihilation cross section and the DM mass, we further
obtain  $C_{DMl}^{ij}(\left< \sigma_{\rm{A}}v\right>, m_\chi)$ which
can be compared with the no-DM hypothesis as:
\begin{eqnarray}\nonumber
    \chi^2(\left< \sigma_{\rm{A}}v\right>,m_\chi)  =
    \phantom{\chi^2(\left< \sigma_{\rm{A}}v\right>,m_\chi)}
     & &  \\ \label{chi2full}
     \Sigma_{l' i j, m n}  \left(C_{EGBl'}^{ij}-C_{DMl'}^{ij}\right) \left( \Delta
    C_{EGBl'}^{ij, mn}\right)^{-1}
    \left(C_{EGBl'}^{mn}-C_{DMl'}^{mn}\right)
\end{eqnarray}
where $\left( \Delta C_{EGBl'}^{ij, mn}\right)^{-1}$ is the inverse
of the covariance matrix. As a first step, however, we might want to
consider a simple expression, for example considering as observables
only the auto-correlation spectra $C_{EGBl'}^{ii}$. Unfortunately, the
naive expression
\begin{equation}  \label{chi2naive}
    \chi^2(\left< \sigma_{\rm{A}}v\right>,m_\chi)=
    \Sigma_{l' i}  \left(C_{EGBl'}^{ii}-C_{DMl'}^{ii}\right)^2 / \left( \Delta
    C_{EGBl'}^{ii,ii}\right)
\end{equation}
turns out to be incorrect because the two spectra $C_{EGBl'}^{ii}$
and $C_{EGBl'}^{mm}$ are correlated, i.e. $\Delta
C_{EGBl'}^{ii, mm}\neq 0$. The correlation, indeed, is given by $\Delta C_{EGBl'}^{ii,
mm}=2/(2l+1)(C_{EGBl'}^{im})^2$, i.e. by the cross correlation between the bands $i$ and $m$, as intuitively expected.
 The variance of $C_{EGBl'}^{ii}$
can be also easily calculated and gives $\Delta C_{EGBl'}^{ii,
ii}=2/(2l+1)(C_{EGBl'}^{ii}+C_N)^2$, where $C_N=4\pi/N_{\rm events}$
is the poisson shot noise. The correct expression is then
\begin{eqnarray}\nonumber
    \chi^2(\left< \sigma_{\rm{A}}v\right>,m_\chi)  =
    \phantom{\chi^2(\left< \sigma_{\rm{A}}v\right>,m_\chi)}
     & &  \\ \label{chi2nocorr}
     \Sigma_{l' ii, mm}  \left(C_{EGBl'}^{ii}-C_{DMl'}^{ii}\right) \left( \Delta
    C_{EGBl'}^{ii, mm}\right)^{-1}
    \left(C_{EGBl'}^{mm}-C_{DMl'}^{mm}\right)
\end{eqnarray}
where $\left( \Delta C_{EGBl'}^{ii, mm}\right)^{-1}$ is the inverse
of $\Delta C_{EGBl'}^{ii, mm}$ and
\begin{eqnarray}\label{autocorrerror}
    \Delta C_{EGBl'}^{ii, ii} = (\delta C_{EGBl'}^{ii})^2 \!\!\!\! & = & \!\!\!\! \frac{2}{2l+1}(C_{EGBl'}^{ii}+C_N)^2
      \\
    \Delta C_{EGBl'}^{ii, mm} = (\delta C_{EGBl'}^{ii}\delta C_{EGBl'}^{mm}) \!\!\!\! & = & \!\!\!\! \frac{2}{2l+1}(C_{EGBl'}^{im})^2
\end{eqnarray}
To take into account  multipole binning and
partial sky coverage, in the above expressions we have to substitute
\begin{eqnarray}\nonumber
    \frac{2}{2l+1} & \rightarrow &  \frac{2}{f_{\rm sky}(2l+1)\Delta l}
      \\
    C_N  & \rightarrow &  \left( \frac{\delta N_{\rm counts}}{\delta \Omega }\right)^{-1} f_{\rm sky}
\end{eqnarray}
where $\Delta l$ is the number of binned multipole in the band l',
and $\delta N_{\rm counts}/ \delta \Omega$ is the density of the
counts per steradian. Notice that in the above formulae no
corrections for the angular resolution (the $\exp(l^2\sigma_b)$
term) are required since we are referring to raw, PSF uncorrected
spectra. Finally, we  disregard the information at $l<10$ to minimize
the effects of the exposure. Let us also notice that, after all, the
value of $\chi^2$ calculated with Eq.\ref{chi2naive}
turns out to be generally similar to the one of Eq.\ref{chi2nocorr}.
This behavior may be due to the different levels
of noise in the various energy bands and in the autocorrelation
spectra which makes the off-diagonal components in the covariance matrix
(which do not have shot noise) generally subdominant.

In case we want to exploit  all the available information, i.e.
including both the auto-correlations and the cross-correlations as
observables, the full expression Eq.\ref{chi2full} for the $\chi^2$
has to be employed. In this case the full covariance matrix becomes
more complicated and we give the corresponding expressions in
appendix B.

Some further comments are in order regarding the calculation of
$\chi^2$. We use always the same set of DM anisotropy maps to
calculate the $C_l$s for the cosmological DM case, despite the fact
that they in principle depend on
$m_\chi$ (and the annihilation mode) through  Eqs.\ref{gammawindow}  and \ref{angspectra}. Consequently,
a different set of maps should be generated for each value of $m_\chi$ and for each channel. 
Fortunately, this dependence is
weak above about 10 GeV where the DM horizon does not depend much on
the DM annihilation energy spectrum so the effects on the
sensitivity plots above 10 GeV are small, accordingly. To better justify
this approximation we can also make a comparison with the case of DM
annihilation from substructures in the Galactic halo where we use a
very different set of maps. The sensitivities generally change by a
factor of a few in this case (see below). The slight dependence of
the maps on DM model for the
cosmological case is thus likely to be smaller than that.

\begin{figure}
\vspace{-0.0pc}
\begin{center}
\begin{tabular}{c}
\hspace{0.0pc}\includegraphics[width=0.90\columnwidth,angle=0]{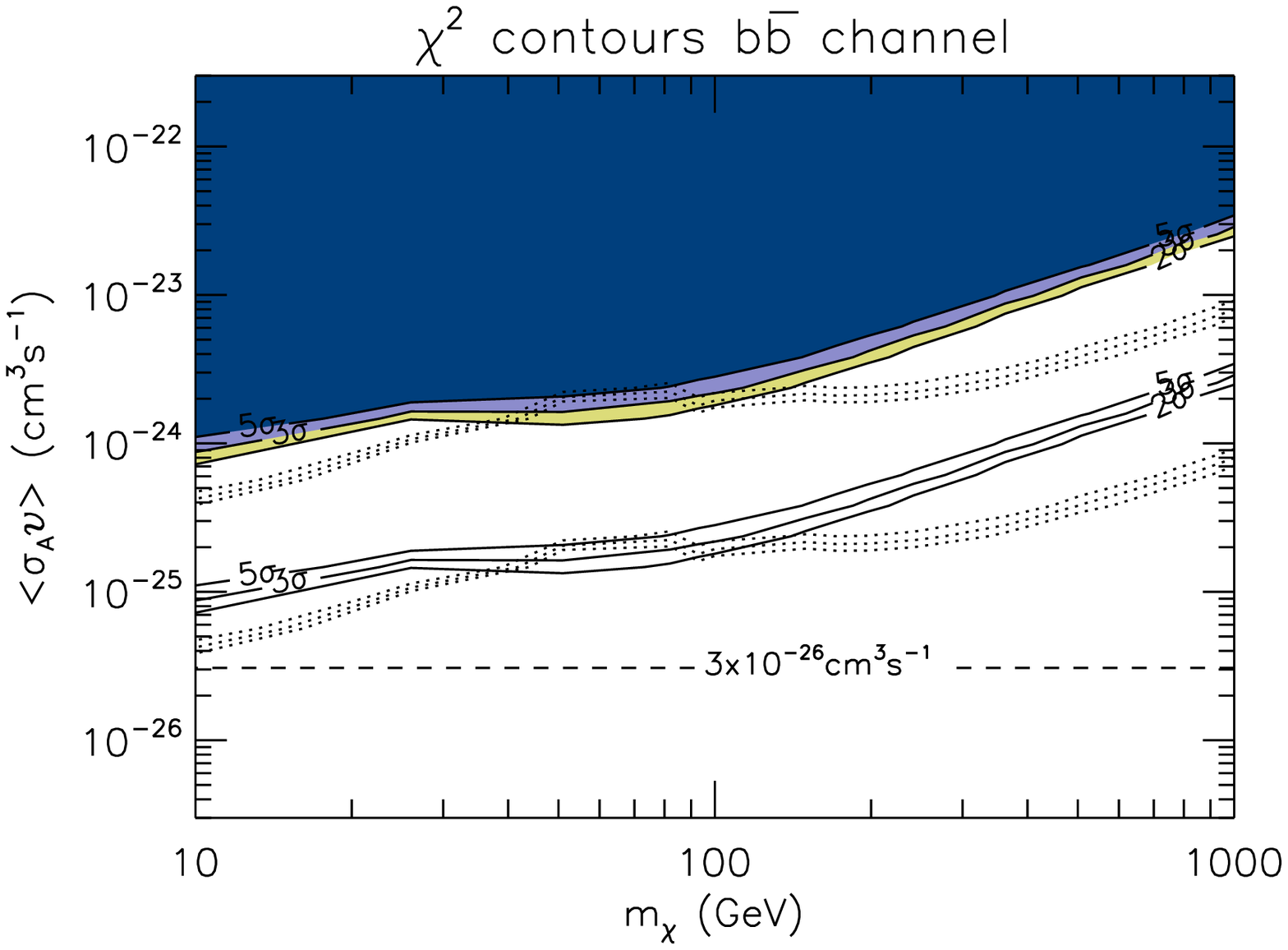}  \\
\vspace{-0.0pc}\includegraphics[width=0.90\columnwidth,angle=0]{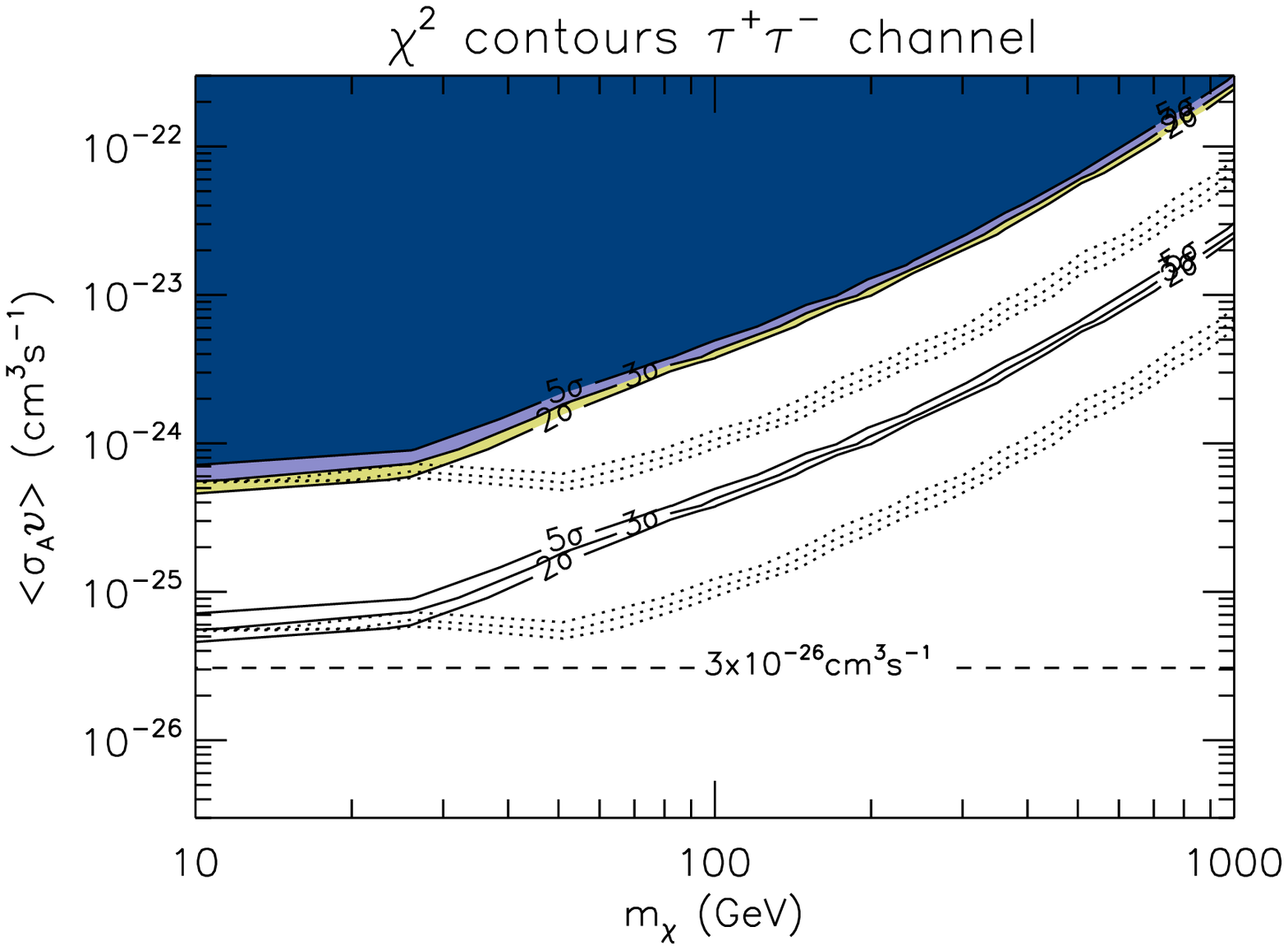} \\
\hspace{0.0pc}\includegraphics[width=0.90\columnwidth,angle=0]{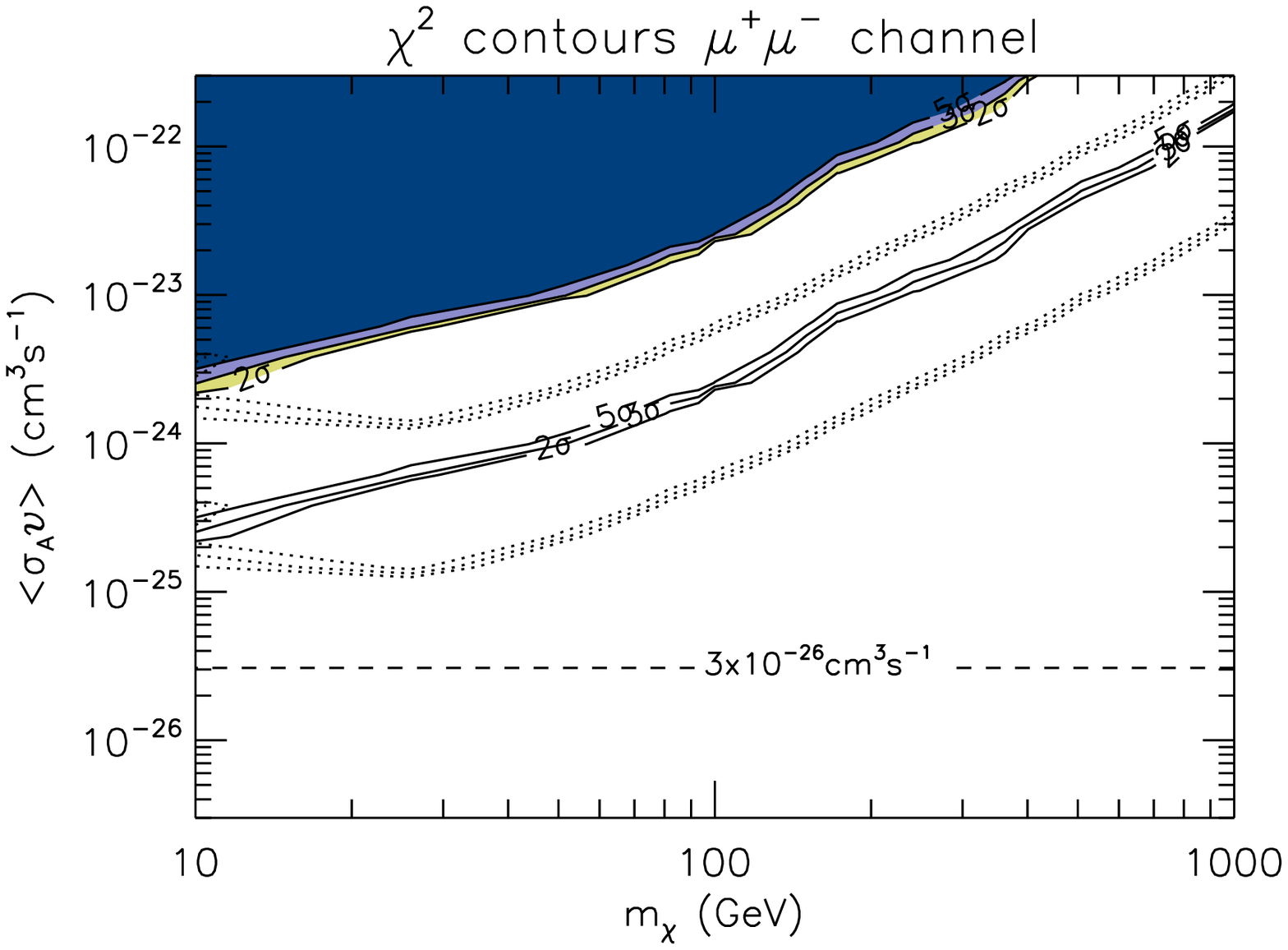}  \\
\end{tabular}
\vspace{-0.5pc}
\end{center}
\hspace{-1.5pc} \caption{Same as previous figure but for the
Galactic substructures case. For a direct comparison with the
cosmological case we use for the DM signal the same normalization(s)
employed for the extra-galactic case. The solid curves are the
sensitivities considering the auto-correlation spectra information
alone. The dotted ones are derived considering also the information
contained in the cross-correlation spectra.\label{fig:limitsb}}
\end{figure}

We can see the results in Fig. \ref{fig:limitsa}, where the
sensitivities  are reported for various annihilation channels.
The absolute normalization of the DM spectra are obtained following
the DM haloes clustering model of \cite{Ullio:2002pj} as described
in the introduction. Two versions of the model are shown: a
conservative version with a NFW profile for the haloes and
no substructures which gives a low DM normalization, and thus
weaker constraints,  and a more optimistic version where the DM signal
is enhanced by the presence of substructures in the DM haloes.
In principle, using the recent results of\cite{Zavala:2009zr}
from the Millenium-II simulation, an order of magnitude enhancement
with respect to the above ``optimistic'' case can be
achieved with a correspondingly better sensitivity (see also \cite{Abdo:2010dk}, in particular Fig.1).

The various channels have generally similar behavior. The
$b\bar{b}$ channel produces quite smooth limits. The
$\mu^+\mu^-$ and $\tau^+\tau^-$ sensitivity curves in contrast have
a steeper slope and exhibit more structure. This is due to the fact that
for these channels the gamma emission is concentrated at higher
energies in a narrow peak near the energy corresponding to the DM
mass. The sensitivity is thus more sensitive to the coarse
binning in energy chosen for this analysis. The sensitivity in the
$\tau^+\tau^-$ channel and especially the $\mu^+\mu^-$ channel at
higher DM masses somewhat decreases due to the lower number of
photons per annihilation (and thus lower statistics) available. 

Including the cross-correlation spectra information does not seem
to improve the sensitivity. This probably has to
be ascribed to the fact that the $\chi^2$ for the auto-correlation case
already includes some cross-correlation information in the
off-diagonal terms of the covariance matrix. The level of
anisotropies is also very low at low energy for this case (see
Fig.\ref{fig:anieni}) and this somewhat reduces the importance of
the information contained in the cross-correlation. The situation is
different for the Galactic DM scenario as discussed below.

We also checked the case in which only the highest energy bands are
employed for the analysis. We would expect an improvement in
sensitivity given that very little DM signal is expected in the
0.4-0.8 GeV band and thus scanning in this energy range introduces a
statistical penalty factor. We find that the sensitivity indeed
improves, but only marginally. On the other hand, the lowest energy
band is useful because being mostly clean of DM it represents a good
calibration for the background  anisotropies.

In Fig. \ref{fig:limitsb} we show  the case in which the DM
anisotropy pattern does not vary with energy. This case is supposed
to mimic the DM emission from unresolved substructures in the Milky
Way. We thus simulate a single anisotropic map which we use for all
energies. The angular spectrum for DM employed to simulate the map
is normalized as in the case of substructures with a minimum mass of
10 $M_\odot$ as derived in \cite{SiegalGaskins:2008ge} and it
roughly corresponds to the extragalactic anisotropies for an energy
of about 300 GeV (see section 2). For a direct comparison with the
cosmological case we use for the Galactic signal  the same
normalization employed for the extragalactic case, which is justified
by the fact that more physically motivated models for the DM emission in
the MW have a large degree of freedom \cite{Ando:2009fp} and
can easily be tuned to match the extragalactic case.

The inclusion of the information from the cross-correlation spectra
in this case, contrary to the cosmological one, gives a significant
improvement, especially for higher DM masses. The reason for this
difference is likely to be found in the very different set of anisotropy
maps employed for the Galactic substructures case. In particular,
the very high degree of anisotropy present also at low energy for
this case  enhances the importance of a direct measurement
of the cross-correlation between low and high energy bands.
In contrast, the very low DM anisotropy at low energy for the
cosmological case (lower than the astrophysical EGB anisotropy)
makes this information almost irrelevant.  We also find in this
case that excluding the lowest energy band
does not  improve the sensitivity significantly.

The sensitivity increases substantially with respect to the cosmological case, especially for high DM
masses. The slope of the
sensitivity curves is indeed slightly less steep. Assuming higher
levels of anisotropy for the Galactic case, which are possible,
\cite{Ando:2009fp,SiegalGaskins:2008ge,SiegalGaskins:2009ux} a
correspondingly higher sensitivity could be achieved. However, the
scaling is not linear. It is worth noting for example that although
at about 10 GeV the Galactic anisotropies are about 2 orders of
magnitudes larger than the extragalactic ones, the sensitivity is
only a factor 2-3 better. The scaling however is better for
$m_\chi\agt100$ GeV where the difference in sensitivity is roughly a
factor of 10.  This non linear scaling is due to the limited limited amount of detected photons (as it is the case at the edges of the sensitivity curve). Here, it is very difficult to distinguish between the two (galactic and extra-galactic) levels of anisotropies even if the latter is a factor 100 larger, since the statistical error bars are also very large. The much larger galactic anisotropy thus only modestly decrease the minimum amount of events required to detect the signal.

\begin{figure}[t]
\vspace{0.0pc}
\begin{center}
\vspace{-0.5pc}\includegraphics[width=0.99\columnwidth,angle=0]{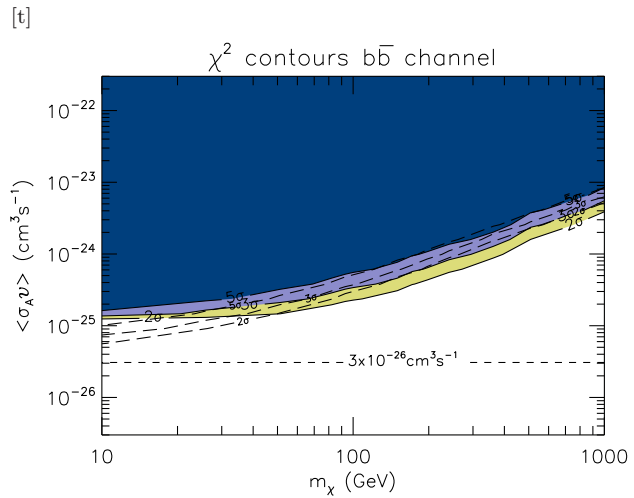}
\vspace{-2.0pc}
\end{center}
\hspace{-1.5pc} \caption{DM sensitivity for annihilation into into $b\bar{b}$,  extragalactic anisotropies and  NFW Halo profile with
substructures, for the two cases of EGB background anisotropies from the correlation term only (same case of Fig.\ref{fig:limitsb}, solid contours) and from a pure  Poisson term only (dashed contours).\label{fig:limitCp}}
\end{figure}

Finally in Fig.  \ref{fig:limitCp} we show the impact on the DM sensitivity of changing the underlying astrophysical
EGB model to the Poisson dominated one. For illustration, only the extragalactic $\tau^+\tau^-$ case with NFW profile and substructures is shown.
The interesting results (perhaps somewhat counter intuitive) is that even with this increased level of background
in the anisotropy spectrum, still the sensitivity to the DM component remains roughly unchanged. The reason is that a
Poisson noise at the level of 5.0e-5 is anyway clearly detected with a statistics of 5 years, so the DM component can still be easily detected as a modulation (as usual, in multipole and in energy) of this dominant Poisson noise anisotropy spectrum. The underlying assumption here, however, is that this new background is fully under control and well characterized. The extent to which this will be true is however not clear and a better understanding may require dedicated simulations and analyses which are beyond the scope of this paper.

\section{Summary and Conclusions}
In the present analysis we have calculated the sensitivity to Dark
Matter annihilation exploiting the characteristic anisotropy
properties that DM would imprint in the Extra-Galactic Gamma
Background. We performed a realistic simulation of the Fermi-LAT
including the effect of the Point Spread Function, effective area
and of the anisotropic sky exposure.

More importantly, we investigated the impact of Galactic
foregrounds on the DM anisotropy sensitivity using a GALPROP model
to produce the sky maps of the Galactic gamma emission. The
contribution of Galactic foregrounds is generally non negligible
even above Galactic latitudes of about $60^\circ$.
Foreground cleaning will thus be an important
(and likely challenging) issue in the analysis of the real data.
For the present work we do not attempt foreground
cleaning (except for masking the Galactic plane and point sources)
and we only require that the anisotropy from DM has to
exceed the level of foregrounds anisotropy to be detectable.
Further, we only analyze a high Galactic latitude region where the
model dependence of the foreground is expected to be moderate. In
this respect our analysis can be regarded as a conservative one: In
reality, if foreground cleaning can be effectively implemented, a
sensitivity to a lower level of anisotropy can be achievable. Also,
a larger fraction of the sky can be employed than the
small conservative one used in the present analysis. We  leave
a more careful investigation of the issue for future work.

Despite the difficulties involved in the foreground analysis, we
find  that the prospects for detecting DM anisotropy signatures are
promising. For some WIMP masses, sensitivity to  $\left<
\sigma_{\rm{A}}v\right> \sim 3\times10^{-26}$cm$^{3}$s$^{-1}$(i.e. the thermal
relic value) can be achieved using the anisotropy information alone.
Interestingly, these constraints are comparable to what is achievable
using the energy spectrum alone \cite{Abdo:2010dk} .
The anisotropy information is thus complementary to the energy
spectrum and provides a cross-check and an increase in the overall
sensitivity to the DM signal.

We also investigated the recently proposed scenario of a population
of DM substructures in the Milky Way halo as predicted by N-body
simulations. These subhaloes give rise to an  annihilation signal
which is almost uniform in the sky and which can be comparable in intensity to
the observed isotropic gamma background. For this case,  the anisotropy
signal is predicted to be even higher than for the extragalactic case
due to the huge clumpiness of the DM distribution that
correspondingly gives an enhanced sensitivity. We find, indeed, that when
increasing the anisotropy the sensitivity to the DM signal increases
as well with an improvement of a factor of $\sim$10 for $m_\chi\agt100$ GeV.

In summary, simulations indicate that ,with the sensitivity of the
Fermi-LAT, detailed anisotropies studies of the gamma-ray sky can
be performed. The next obvious step will be the analysis of the actual data.  
Anisotropy analyses will soon provide a
complementary source of information which will give further insights into the
DM problem and the properties of high-energy astrophysical sources.

\section*{Acknowledgments}
The authors would like to thank Toby Burnett, Jennifer Siegal-Gaskins and Eiichiro Komatsu 
for their numerous comments and suggestions for improvement,  
and Chris Savage for a carefully reading of the manuscript. 
We also wish to thank Gianfranco Bertone for  organizing the workshop ``Gamma in Zurich''  which stimulated 
many interesting discussions on gamma-ray anisotropies.

The \textit{Fermi} LAT Collaboration acknowledges generous ongoing support
from a number of agencies and institutes that have supported both the
development and the operation of the LAT as well as scientific data analysis.
These include the National Aeronautics and Space Administration and the
Department of Energy in the United States, the Commissariat \`a l'Energie Atomique
and the Centre National de la Recherche Scientifique / Institut National de Physique
Nucl\'eaire et de Physique des Particules in France, the Agenzia Spaziale Italiana
and the Istituto Nazionale di Fisica Nucleare in Italy, the Ministry of Education,
Culture, Sports, Science and Technology (MEXT), High Energy Accelerator Research
Organization (KEK) and Japan Aerospace Exploration Agency (JAXA) in Japan, and
the K.~A.~Wallenberg Foundation, the Swedish Research Council and the
Swedish National Space Board in Sweden.

Additional support for science analysis during the operations phase is gratefully
acknowledged from the Istituto Nazionale di Astrofisica in Italy and the Centre National d'\'Etudes Spatiales in France.



\appendix

\section{Angular power spectrum}
We summarize in this appendix the conventions and notation used for
the anisotropy analysis. The anisotropy of a signal on the sky can
be quantified in terms of the power spectrum i.e. the spherical
harmonic transform on the sphere. More specifically, a map
on the sphere $f(\hat{\Omega})$  can be decomposed in spherical
harmonics as $f(\hat{\Omega})= \Sigma_{lm} a_{lm}
Y^l_m(\hat{\Omega})$  and the angular power spectrum can then be
defined as
\begin{equation}
  C_l \equiv <|a_{lm}|^2>
\end{equation}
where $<...>$ indicates the statistical ensemble average. The
quantity
\begin{equation}\label{pspectrum}
  \hat{C_l}= \sum_m \frac{|a_{lm}|^2}{2 l+1}
\end{equation}
is an unbiased estimator of the true power spectrum $C_l$, i.e.
$<\hat{C_l}>=C_l$.

In the case of two maps, besides the two autocorrelation power
spectra $C_{l}^1$ and $C_l^2$, further information is contained in
the \emph{cross-correlation power spectrum} which can be estimated as
\begin{equation}\label{cspectrum}
  \hat{C_l}^{12}=\sum_m
\frac{a_{lm}^{1}a_{lm}^{2*}}{2l+1} = \sum_m
\frac{a_{lm}^{1*}a_{lm}^{2}}{2l+1} 
\end{equation}
where the equivalence of the two expressions comes from  the equality: $a_{lm}^{i} = (-1)^m a^{i*}_{l-m}$, $i=1,2$. 
The three spectra are of course not independent but are part of a
covariance matrix:
\begin{equation}
\left(
    \begin{array}{cc}
        C_l^{1}& C_l^{12} \\
       C_l^{12} & C_l^{2}
     \end{array}
\right)
\end{equation}
which has to be taken into account when evaluating the errors for
the forecast. This is described in more details in the next section.

In practice, to calculate the power spectra, the harmonic transform
$a_{lm}$ is performed with HEALpix \cite{Gorski:2004by} after
binning the events into maps of the HEALpix format itself. Due to
the finite number $N$ of collected events beside the intrinsic
anisotropies the data also contain noise (shot noise or white noise)
which appears in the power spectrum as a constant whose value is
$4\pi/N$. The noise sets the limit above which the intrinsic
anisotropies can be detected. It is standard to remove the shot
noise from the spectrum showing only the intrinsic power spectrum.
It is also customary to show not the power spectrum $C_l$ but the
quantity $l(l+1)C_l/2\pi$.

\section{$\chi^2$ covariance matrix}
In the general case in which we include as observables the
cross-correlation power spectra the covariance matrix can be
calculated by propagation of the errors from the expressions
\ref{pspectrum} and \ref{cspectrum} and then taking the ensemble
average. The variance $\Delta C_{EGBl'}^{ij, ij}$ of the
cross-correlation spectrum $C_{EGBl'}^{ij}$ ($i\ne j$) can be written as
\begin{eqnarray}\nonumber
    \Delta C_{EGBl'}^{ij, ij}  =  <(\delta C_{EGBl'}^{ij})^2> \;\;\;\;\;\;\;\;\;\;\;\;\;\;\; & = &
    \\ \label{crosscorrerror}
    \frac{1}{2l+1}\left((C_{EGBl'}^{ij})^2+(C_{EGBl'}^{ii}+C_{N_i})(C_{EGBl'}^{jj}+C_{N_j})
    \right) & &
\end{eqnarray}
We then have the case of covariance $\Delta C_{EGBl'}^{ii, mn}$
between an auto-correlation spectrum $C_{EGBl'}^{ii}$ and
cross-correlation spectrum $C_{EGBl'}^{mn}$. For this case we have
\begin{eqnarray}\nonumber
    \Delta C_{EGBl'}^{ii, mn}  =  <\delta C_{EGBl'}^{ii}\delta C_{EGBl'}^{mn}> \;\;\;\;\;\;\; & = &  \\
    \frac{2}{2l+1}\left((C_{EGBl'}^{ii}+C_{N_i})(C_{EGBl'}^{mn})
    \right) & &
\end{eqnarray}
For the case of covariance between two cross-correlation spectra we
have to distinguish two cases. In the case of covariance $\Delta
C_{EGBl'}^{ij, jn}$ between $C_{EGBl'}^{ij}$ and $C_{EGBl'}^{jn}$
where there is an energy band (an index) in common we have:
\begin{eqnarray}\nonumber
    \Delta C_{EGBl'}^{ij, jn}  =  <\delta C_{EGBl'}^{ij}\delta C_{EGBl'}^{jn}> \;\;\;\;\;\;\;  & = &  \\
    \frac{1}{2l+1}\left((C_{EGBl'}^{jj}+C_{N_j})(C_{EGBl'}^{in})+(C_{EGBl'}^{ij})(C_{EGBl'}^{jn})
    \right) & &
\end{eqnarray}
If instead there is no energy band (no index) in common $\Delta
C_{EGBl'}^{ij, mn}$ finally we have
\begin{eqnarray}\nonumber
    \Delta C_{EGBl'}^{ij, mn}  =  <\delta C_{EGBl'}^{ij}\delta C_{EGBl'}^{mn}> \;\;\;\;\;\;\;  & = &  \\
    \frac{1}{2l+1}\left((C_{EGBl'}^{im})(C_{EGBl'}^{jn})+(C_{EGBl'}^{in})(C_{EGBl'}^{jm})
    \right) & &
\end{eqnarray}

Expression (\ref{autocorrerror}) and (\ref{crosscorrerror}) can also
be rewritten in the more familiar form
\begin{equation} \label{gausserrors1}
  \frac{\delta C_l}{C_l} =
  \sqrt{\frac{2(1+{C_N}/C_l)^2}{(2l+1)\Delta l f_{\rm sky}}},
\end{equation}
for the auto-correlation spectra and by
\begin{eqnarray}\nonumber\label{gausserrors2}
  && \!\!\!\!\!\!\!\!\!\!\!\! \frac{\delta C_l^{{12}}}{C_l^{{12}}} =
  \sqrt{\frac{1}{(2l+1)\Delta l f_{\rm sky}}} \,\, \times   \\
  && \!\!\!\!\!\!\!\!\!\!\!\! \sqrt{\left( 1+      \frac{ C_l^{1}\; C_l^{2}}{C_l^{{12}} \; C_l^{{12}}}
  \left(1+{C_{N_1}}/C_l^{1}) (1+{C_{N_2}}/C_l^{2}
  \right)    \right) }   ,
\end{eqnarray}
for the cross-correlation spectra, where again
\begin{eqnarray}
    C_{N_i}  = \left( \frac{\delta N^i_{\rm counts}}{\delta \Omega }\right)^{-1} f_{\rm sky}
\end{eqnarray}
and  $\Delta l$ is the number of binned multipole in the band $l'$,
and $\delta N^i_{\rm counts}/ \delta \Omega$ is the density of the
counts per steradian in the map $i$.

\end{document}